%% file: bare_jrnl.tex
\documentclass[journal]{IEEEtran}
\usepackage{amsmath,amsfonts}
\usepackage{algorithmic}
\usepackage{algorithm}
\usepackage{array}
\usepackage[caption=false,font=normalsize,labelfont=sf,textfont=sf]{subfig}
\usepackage{textcomp}
\usepackage{stfloats}
\usepackage{url}
\usepackage{verbatim}
\usepackage{graphicx}
\usepackage{cite}
\usepackage{soul,color}
\usepackage{svg}
\usepackage{tabularray}
\UseTblrLibrary{booktabs}
\usepackage{wrapfig}
\usepackage{tablefootnote}
\usepackage{fancyhdr}

\newcommand{\DM}[1]{\textcolor{black}{#1}}
\newcommand{\LH}[1]{\textcolor{black}{#1}}

\newcommand{\DMRev}[1]{\textcolor{black}{#1}}
\newcommand{\DMMinorRev}[1]{\textcolor{black}{#1}}

\fancypagestyle{firstpage}{%
 \fancyhf{}%
 \fancyhead[C]{This work has been submitted to the IEEE for possible publication. Copyright may be transferred without notice, after which this version may no longer be accessible.}
}
\begin{document}

\title{Waveform Learning under Phase Noise Impairment for Sub-THz Communications
}

\author{
\IEEEauthorblockN{Dileepa~Marasinghe$^{1}$,
        Le Hang~Nguyen$^{2}$, Jafar~Mohammadi$^{2}$, Yejian~Chen$^{2}$,  \\ Thorsten~Wild$^{2}$ and~Nandana Rajatheva$^{1}$}\\
\IEEEauthorblockA{$^{1}$University of Oulu, Finland}\\
\IEEEauthorblockA{$^{2}$Nokia Bell Labs, Stuttgart, Germany }

}




\maketitle
\thispagestyle{firstpage}  

\begin{abstract}
The large untapped spectrum in sub-THz allows for ultra-high throughput communication to realize many seemingly impossible applications in 6G.
Phase noise (PN) is one key hardware impairment, which is accentuated as we increase the frequency and bandwidth. 
Furthermore, the modest output power of the power amplifier demands limits on peak to average power ratio (PAPR) signal design. 
In this work, we design a PN-robust, low PAPR single-carrier (SC) waveform by geometrically shaping the constellation and adapting the pulse shaping filter pair under practical PN modelling and adjacent channel leakage ratio (ACLR) constraints for a given excess bandwidth. We optimize the waveforms under conventional and state-of-the-art PN-aware demappers. Moreover, we introduce a neural-network (NN) demapper enhancing transceiver adaptability.
We formulate the waveform optimization problem in its augmented Lagrangian form and use a back-propagation-inspired technique to obtain a design that is numerically robust to PN, while adhering to PAPR and ACLR constraints.
\DM{The results substantiate the efficacy of the method, yielding up to 2.5 dB in the required $E_b/N_0$ under stronger PN along with a PAPR reduction of 0.5 dB. Moreover, PAPR reductions up to 1.2 dB are possible with competitive BLER and SE  performance in both low and high PN conditions.}
\end{abstract}



\section{Introduction}
\input{1_introduction}

\section{System model}
\label{3_SystemModel}
\input{3_SystemModel}

\section{Problem formulation}
\label{4_Problemformulation}
\input{4_Problemformulation}

\section{Optimization Methodology}
\label{5_OptimizationMethodology}
\input{5_OptimizationMethodology}

\section{Evaluation Setup}
\label{6_EvaluationSetup}
\input{6_EvaluationSetup}

\section{Results and discussion}
\label{7_Results}
\input{7_Results}

\section{Conclusion}
\label{8_ConculsiosnsAndOutlook}
\input{8_ConculsiosnsAndOutlook}


\bibliographystyle{IEEEtran}
\bibliography{bibi}

\end{document}

%% file: 1_introduction.tex
Scoring the Terabits goal envisioned for 6G appears increasingly tangible by tapping into the large bandwidth available at sub-THz range, i.e., 90-300 GHz. \DMMinorRev{In addition to enabling extreme throughputs, such bandwidths could be leveraged for several purposes: offloading mobile traffic from lower frequency bands, supporting high-throughput fixed wireless access and backhauls, and facilitating high-resolution localization and sensing capabilities envisioned for 6G networks\cite{HEX23-D23}.} However, significant challenges persist in addressing channel conditions and hardware impairments at such high frequencies, which necessitates a rethink of the physical layer processing starting from the waveform design. Realizing the sub-THz links will favor towards single-carrier (SC) transmissions\cite{HexaXSubTHz}, because of lower peak to average power ratio (PAPR) and less sensitivity to hardware impairments over the current new radio (NR) multi-carrier (MC) waveforms. Consequently, SC transmission facilitate enhanced coverage, reliability and most importantly more energy-efficient communications, which is one of the essential requirements for 6G. Among the SC waveform candidates, pure single carrier frequency domain equalization (SC-FDE)\cite{FalconerSCFDE} stands out as a potential 6G waveform  for sub-THz due to the transmitter's simplicity and its compatibility with the incumbent NR waveforms \cite{LeHangWaveform}. 

A key hardware impairment that significantly hinders the links in sub-THz is the \DM{excess} phase noise (PN), which is a result of the imperfect local oscillators. 
\DMMinorRev{Enhancing the spectral purity of the local oscillators at the hardware level incurs significant costs, rendering their widespread adoption impractical. Therefore, addressing phase noise necessitates the development of solutions through meticulous signal and algorithmic design. 
The PN is characterized by the power spectral density (PSD), which describes the noise power per Hz at a certain offset from its operating frequency. 
The sub-THz frequency is generated by up-converting from a reference frequency with multiple stages, where the PN power increases by 6 dB for every doubling of the carrier frequency.} 
The PN PSD can be modelled as a superposition of two random processes: a correlated Wiener 
process and an uncorrelated Gaussian process\cite{CEA_Leti_PN_model}, \DMMinorRev{whose variance is determined by the noise floor of the PSD.} The conventional approach to address the contribution of the low frequency Wiener PN is by sending phase tracking reference signals (PTRS) to track and compensate for the phase errors at the receiver\cite{PN_tracking_Linear}. Arguably, the main motivation for developing sub-THz technology is the abundance of 
the wide bandwidth. \LH{The contribution of Gaussian PN strongly depends on the system bandwidth: the wider the bandwidth, the greater the influence of Gaussian PN. Due to its uncorrelated nature, Gaussian PN cannot be well mitigated using PTRS. }
Thus, even 
after compensation by PTRS, the Gaussian PN remains, which deteriorates the decoding accuracy. 
The distribution of received symbols impacted by PN exhibits a significantly different noise distribution due to the phase error compared to the thermal noise 
distribution considered in the conventional communication systems. Therefore, depending on the symbols' constellation, the waveform exhibits different 
robustness against PN, particularly against the Gaussian PN part. 

The inherent non-linear nature of the power amplifier (PA) imposes limitations on its output power and introduces distortion to the waveform \cite{Hardy_PA}. 
A waveform with low PAPR enables operation with a smaller back-off, allowing the PA to function closer to its optimal point, thus saving considerable amount of energy, while increasing the coverage range.
Therefore, an important characteristic of the transmit waveform to be considered in the sub-THz regime is the PAPR of the waveform \cite{LeHangWaveform}. Although SC waveform results in lower PAPR compared to its MC counterparts, further PAPR reduction is attainable by
the constellation geometry and the impulse response of the band-limiting transmit filter used at the baseband processing level. 
Simultaneously, the adjacent channel leakage ratio (ACLR) of the waveform, determined by the power spectral density of the transmit filter should be kept under a certain limit imposed by regulatory bodies for radio communications to avoid interference to adjacent channels.
%
\subsection{Prior Art}
In this subsection, we summarize the related literature for PN-robust constellation designs and waveform learning approaches.
An algorithm to select constellation points on a lattice, which minimizes the symbol error probability under a Tikhonov PN characterization has been presented in \cite{Foschini}. Most of the prior works consider the residual PN, characterized with a Gaussian process assuming an ideal tracker/estimator, which removes the correlated part of the PN \cite{KrishnanPNOptimization}. The article \cite{KrishnanSoftMetrics} investigates the maximum likelihood detection in the presence of strong phase noise. The authors derive soft metrics under the assumptions of Gaussian PN and when the probability density function (PDF) of the phase error is unknown but its central moments are known. The subsequent derivation of the symbol error probability encompasses arbitrary constellations. Further contributions in \cite{KrishnanPNOptimization} introduce two approximate maximum likelihood detectors, predicated on high \DM{signal-to-noise ratio} (SNR) and low PN assumptions. These detectors are harnessed for constellation designs via gradient-based optimization, targeting maximum mutual information and symbol error probability, which shows that optimally designed constellations prove resilience against strong phase noise compared to the conventional constellations. However, the optimization criteria does not account for PAPR. Spiral constellations presented in \cite{Spiral} provided a semi-analytic method for defining the constellation, while employing a detection rule originally proposed in \cite{KrishnanPNOptimization}. Recent works aiming at sub-THz on the optimum demodulation and constellation design are presented in \cite{CEATLetiOptimumDemodulation} and \cite{CEALetiDigitalDesign}, which proposed polar quadrature amplitude modulation (QAM) and a method of soft-decision decoding compatible with channel coding based on a detection rule proposed in \cite{KrishnanPNOptimization}. However, as noted in \cite{CEALetiDigitalDesign}, both spiral constellations and the polar-QAM have similar PAPR characteristics and their performance improvements under PN come at a significant deterioration of the PAPR. \DM{The work in \cite{NortheasternPN} shows the importance of jointly assessing PN and PAPR in the context of sub-THz communications under practical PN models. The authors first measure PN characteristics of a practical setup, then tune the PN models recommended by 3rd generation partnership project (3GPP) \DM{\cite[\S6.1.10]{3gpp.38.803}}, to their measurements and evaluate the performance of numerous waveforms under PN and a PAPR penalty. However, they do not explicitly design a waveform tailored to sub-THz and the thermal noise floor of the PN model is not considered. Furthermore, the pulse shaping filters are not accounted in the evaluation.}

\DMMinorRev{Traditionally in the context of optimizing the waveform, the analysis of these transceiver components is carried out individually or in sub-groups, subjected to different optimization criteria. This approach is necessitated by computational limitations and the convenience of tractability. Recently, with the growth of computational capability for developing artificial intelligence and machine learning (AI/ML) methods, an emerging line of work is to model transceivers as autoencoders with neural networks \cite{SebastianTrainableComm}. Optimization or ``learning" is then carried out in an end-to-end fashion, to maximize the information rate, facilitated by large-scale gradient backpropagation. In \cite{DeepWaveform}, authors
propose a deep complex-valued convolutional network, which is able to recover bits from the time-domain orthogonal frequency division multiplexing (OFDM) signals without relying on any explicit fourier transform operations. The authors in  \cite{FaycalPilotless} showcased pilotless communications for  OFDM, by training the transceiver chain in an end-to-end fashion employing a trainable constellation and a deep neural receiver based on convolutional neural networks (CNN) \cite{DeepRX}. Extending this idea, the potential of learning a waveform  under a non-linear power amplifier was demonstrated in \cite{DaniHexaXPA}, by shaping the constellation and introducing a CNN at the transmitter to contain the out-of-band radiation resulting from the non-linearity of the power amplifier. In \cite{MathieuConference}, an OFDM waveform learning problem under PAPR and ACLR constraints was investigated where the transmitter and receiver are both modelled as CNNs.}  
The investigation in \cite{WaveformFaycal} tackles end-to-end learning of SC waveforms under PAPR and ACLR constraints. Leveraging a deep convolutional neural receiver, the study optimizes constellations, bit labeling, and pulse shaping filters for SC waveforms across AWGN and fading channels. Noteworthy gains in data rate are achieved with reduced PAPR and ACLR compared to conventional waveforms. However, it is crucial to note that the pulse shaping filters in this approach are \DM{learned in the frequency-domain, and impulse response is obtained through the inverse Fourier transform. This leads to} complex-valued \DM{time-domain filters, which are} asymmetrical around the peak, diverging from conventional filters. 
Additionally, the deep neural receiver incurs a high computational cost.

In contrast to these, we model a practical PN condition, which includes the Wiener part in this work and aim to design or ``learn" a waveform robust to the residual Gaussian PN. Furthermore, we model the oversampled channel compared to the symbol time channel model in \cite{WaveformFaycal} and aim to learn real-valued filters directly with the impulse responses. Moreover, we consider conventional processing, which has low complexity compared to a deep neural receiver.

\subsection{Our contribution}
The main contributions of this paper are the following:
    \begin{itemize}

    \item \DMMinorRev{We introduce an end-to-end SC transceiver model optimized for sub-THz communication, incorporating practical PN characterization and compensation. The waveform optimization problem is formulated with trainable constellations and pulse shaping filters, constrained by PAPR and ACLR for the filter within a specified excess bandwidth.}
    \item \DMMinorRev{We propose the utilization of a neural network (NN) demapper as a replacement for analytically derived demappers, enhancing the trainability and optimization of signal detection within the transceiver.}
    \item \DMMinorRev{Through data-driven optimization, we demonstrate the feasibility of learning robust waveforms for the SC transceiver under the conventional Additive White Gaussian Noise (AWGN) optimal demapper, state-of-the-art PN-robust demappers, and the proposed NN demapper, with enhanced resilience against PN, while simultaneously achieving lower PAPR under specified ACLR and excess bandwidth for the filter.}
    \item \DMMinorRev{We conduct an extensive analysis of the characteristics of the learned constellations and filters, and quantify the performance gains in terms of PAPR, block error rate (BLER) and spectral efficiency (SE) over the baselines.} 
\end{itemize}


%% file: 3_SystemModel.tex
\begin{figure*}[ht]
    \centering
    \includegraphics[width=\textwidth]{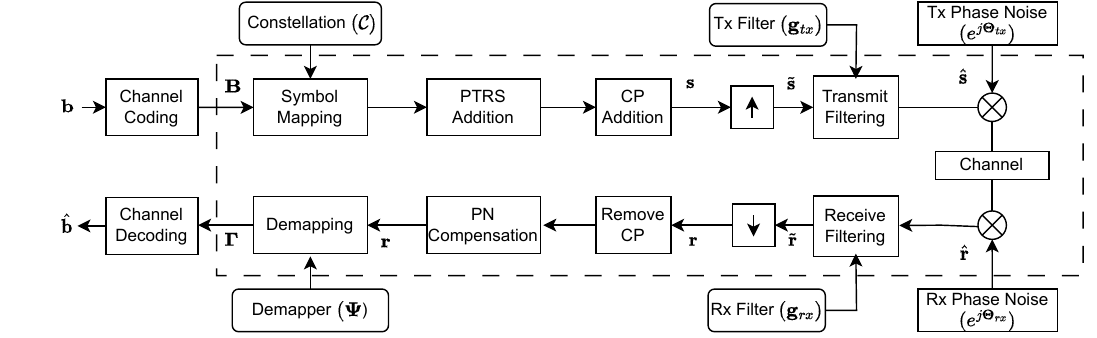}
    \caption{End-to-end system model for the \DM{SC} transceiver} 
    \label{fig:ened_to_end_model}
\end{figure*}

In this section, we introduce the discrete-time system model considered in this work. We consider a conventional \DM{SC} transceiver as shown in Figure \ref{fig:ened_to_end_model}, which is impaired by phase noise in the transmit and receive sides.  
Let $\mathbf{b}$ be the vector of data bits that are to be transmitted through the system. First, $\mathbf{b}$ is channel coded with rate $r<1$ to a coded matrix of bits ${\mathbf{B}}_{N_D \times K}$, where $K$ is the number of bits per symbol and $N_D$ is the number of modulated data symbols. 
Then, each column of $K$ bits are mapped to the corresponding symbol based on a constellation $\mathcal{C} \in \mathbb{C}^{2^K} $ with the respective bit-labelling following the \DM{well-known} bit-interleaved coded modulation (BICM) scheme\DMMinorRev{\cite{BICM}}. 
Next, PTRS time samples are organized into $Q$ groups and evenly distributed between the mapped data symbols, with each group containing $N_P$ time samples.
They are primarily intended for compensating the Wiener part of the PN at the receiver. This is followed by an $N_{CP}$-cyclic prefix (CP) extension to the block to form the transmit symbol vector $\mathbf{s} \in \mathbb{C}^{N}$, where $N = N_D + Q \times N_{P} + N_{CP}$, is the length of the transmission block. Then $\mathbf{s}(n)$ is upsampled with an oversampling factor of $M$ compared to the symbol rate such that, 
\begin{equation}
\Tilde{\mathbf{s}}(n)=\begin{cases} \begin{aligned} 
    &\mathbf{s}\left(\frac{n}{M}\right), \quad \frac{n}{M} \in \mathbb{Z},\\
    & 0, \quad\quad\quad\quad \text{otherwise.}
\end{aligned} \end{cases}
\end{equation}
The upsampled signal $\Tilde{\mathbf{s}}(n)$ is filtered by the transmit filter $\mathbf{g}_{tx}$ as,
\DM{\begin{align}
    \hat{\mathbf{s}}(n) &= \Tilde{\mathbf{s}}(n) \ast \mathbf{g}_{tx}(n),\\
                        &= \sum_k \mathbf{s}\left(\frac{n-k}{M}\right)\mathbf{g}_{tx}(k),
\end{align}}
where $\ast$ denotes the linear convolution. The filtered signal $\mathbf{\hat{s}}(n)$ is then sent to the analog front end (AFE) of the transmitter, where the phase jitter of the local oscillator impairs the signal with PN samples $\mathbf{\Theta}_{tx}(n)$. Then the PN impaired transmit signal goes though the channel. 
\DM{In case of our targeted SC-FDE deployment with the sub-THz channel, additional pilots in the transmitter and channel estimation in the receiver followed by an equalizer are utilized. Since the focus is on the PN impairment, we emulate a scenario of perfect channel equalization by deploying an AWGN channel.} The received signal at the AFE of the receiver is further impaired by the receiver phase noise samples $\mathbf{\Theta}_{rx}(n)$. The sampled received signal $ \hat{\mathbf{r}}(n)$ can be written as, 
\DM{\begin{align}
     \hat{\mathbf{r}}(n) &= \hat{\mathbf{s}}(n) e^{j\mathbf{\Theta}_{tx}(n)} e^{j\mathbf{\Theta}_{rx}(n)} + \mathbf{w}(n),\\
                        &= \sum_k \mathbf{s}\left(\frac{n-k}{M}\right)e^{j\mathbf{\Theta}_{tx}(n)} e^{j\mathbf{\Theta}_{rx}(n)}\mathbf{g}_{tx}(k)  + \mathbf{w}(n),
\end{align}
where $\mathbf{w}(n) \sim \mathcal{CN}(0,\,\sigma^2_n)$, denotes the complex AWGN with variance $\sigma^2_n$. }
The sampled signal is then filtered with the receive filter $\mathbf{g}_{rx}$ as, 
\begin{equation}
    \Tilde{\mathbf{r}}(n) = \hat{\mathbf{r}}(n) \ast \mathbf{g}_{rx}(n)
\end{equation}
and downsampled by a factor of $M$, which provides the received symbols,
\DM{\begin{multline}
\mathbf{r}(n) = \sum_l \sum_k \mathbf{s}\left(\frac{l-k}{M}\right)e^{j\mathbf{\Theta}_{tx}(l)} e^{j\mathbf{\Theta}_{rx}(l)}\mathbf{g}_{tx}(k) \\ \mathbf{g}_{rx}(nM-l)  + \mathbf{\Tilde{w}}(n),
\end{multline}
where  $\mathbf{\Tilde{w}}(n)$ has the same statistics as $\mathbf{w}(n)$.}
First, the CP is removed from $\mathbf{r}(n)$.  The frequency domain equalization step is transparent thanks to the AWGN channel deployment. PTRS time samples are then extracted for PN compensation. We describe this PN compensation in more detail in subsection \ref{phase_noise_compensation}. The resulting data symbol block ${\mathbf{r}}(n)$  is sent to the demapper function $\bf{\Psi}$,  to generate soft estimates $\bf{\Gamma}$, for each bit transmitted from $\mathbf{B}$.  We consider bit-metric decoding (BMD), \DM{in which} the \DMMinorRev{log-likelihood ratio (LLR)} for each bit is calculated as,
\begin{equation}
    \Gamma_{n,k} = \log \left( \frac{\sum_{c_i\in \mathcal{C}(1, k)} p(\mathbf{r}(n)| c_i)}{\sum_{c_i\in \mathcal{C}(0, k)} p(\mathbf{r}(n)|c_i)} \right),
\label{eq_LLR_main}
\end{equation}
where $\Gamma_{n,k}$ is the LLR for the $k^\text{th}$ bit $(0 \leq k \leq K - 1)$ of the $n^\text{th}$ symbol $(0 \leq n \leq N_D - 1)$, and $\mathcal{C}(0, k)$ and $\mathcal{C}(1, k)$ are the subsets of $\mathcal{C}$, which contains all constellation points with the $k^\text{th}$ bit label set to $0$ and $1$ respectively. Since the symbols are equiprobable, $p(r|c)$ denotes the likelihood of $r$ given $c$ was transmitted. Note that the definition of the LLR here is the inverse of the conventional definition, which is intentional to align with the logit definition commonly used in the machine learning frameworks. 
The soft estimates are then sent to the channel decoder for generating the bit estimates, $\bf{\hat{b}}$ corresponding to the transmitted bits.

\subsection{Phase noise models}
The main focus in this work is to design a waveform, which is robust against PN. Hence, accurate modelling of the phase noise is \DM{important}. Furthermore, our approach is data-driven, which allows us to approximate the practical PSD characteristics of the PN to a sufficient accuracy in contrast to models that neglect the Wiener part based on an ideal estimator assumption \cite{KrishnanPNOptimization},\cite{CEALetiDigitalDesign}. Therefore, we consider the well-established multi pole-zero PN model \DMMinorRev{for the transmitter} recommended by \DM{\cite[\S6.1.10]{3gpp.38.803}}. The PSD we consider is given by,
\begin{equation}\label{PN_3GPP_model_10}
S\left( f \right) = \text{PSD0}\frac{\prod\limits_{n = 1}^N 1 + {\left( {\frac{{{f}}}{{{f_{z,n}}}}} \right)^{{\alpha _{z,n}}}}} {\prod_{m=1}^{N} 1+\left(\frac{f}{f_{p, m}}\right) ^{\alpha_{p, m}}},
\end{equation}
where $\text{PSD0}$ is the power spectrum at zero frequency, $f_{z,n}$ and $f_{p,m}$ are zero and pole frequencies. $\alpha_{z,n}$ and $\alpha_{p,m}$ are the powers at each zero and pole frequencies. \DMMinorRev{ The noise power is scaled to the needed carrier frequency ($f_c$) by a factor $20 \log (\frac{f_c}{f_{Ref}})$, where $f_{Ref}$ is the reference frequency of the PSD measurements.}

\begin{figure}[ht]
    \centering
    \includegraphics[width=0.77\columnwidth]{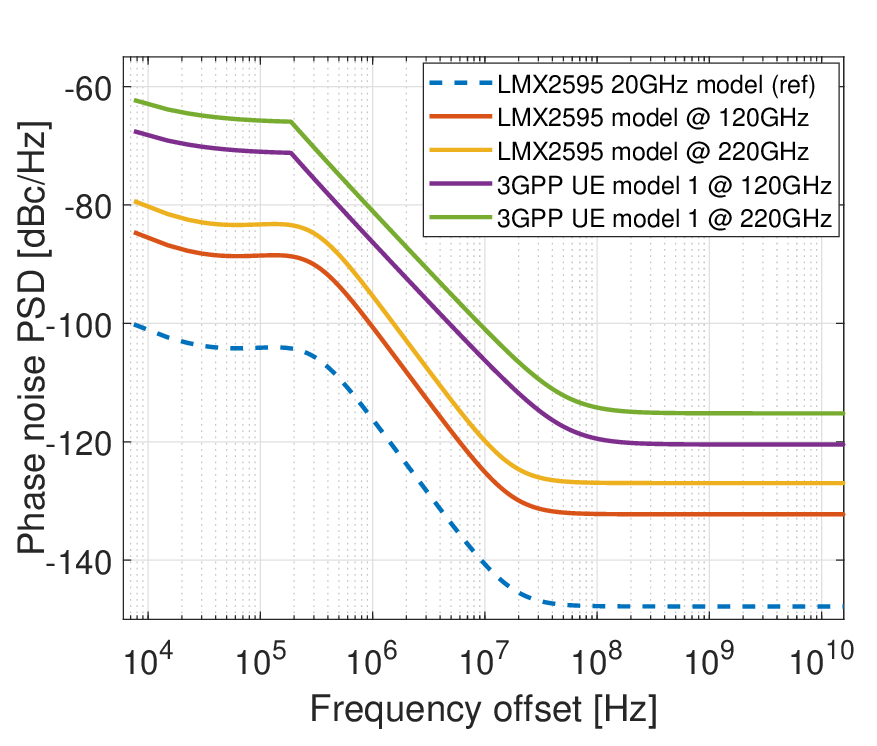}
    \caption{PSDs of the PN models}
    \label{fig:PN_models}
\end{figure}

\DMMinorRev{For the receiver, we consider the model in \cite[\S6.1.11]{3gpp.38.803}, which has a smaller loop bandwidth (LBW), recommended for user equipment due to low cost and less power consumption but with a compromise in phase noise level. This model is specified in log scale compared to the earlier model and is characterized by,
\begin{equation}
S\left( f \right) = \begin{cases}
    S_{Ref}(f) + S_{PLL}(f) & f \leq \text{LBW},\\
    S_{VCO_{v2}}(f) + S_{VCO_{v3}}(f) & f > \text{LBW},
\end{cases}
\end{equation}
where each component added is given by,
\begin{equation}
 S_{Ref/PLL/VCO_{V2}/VCO_{V3}}\left( f \right) = \text{PSD0}\frac{ 1 + {\left( {\frac{{{f}}}{{{f_{z}}}}} \right)^{k}}} { 1+f^k} (dB),
\end{equation}
and the $\text{PSD0}$ as,
\begin{equation}
    \text{PSD0} = \text{FOM} + 20 \log f_c - 10 \log(\frac{\text{POW}}{1mW}) (dB).
\end{equation}
Here, FOM is the figure of merit and POW is the consumed power.
We generate the phase noise samples 
using the filtered Gaussian technique, where initial noise samples are drawn from a standard normal distribution 
and filtered using the PSDs according to the specified models in the frequency domain and then converted back to the time domain. Then the generated phasors are multiplied with the oversampled pulse-shaped signal, which implicitly simulates the intersymbol interference (ISI) caused by the phase noise. PN model the parameters used for evalaution are listed in Table \ref{TxPN_model} and Table \ref{Rx_PN_model} while Figure \ref{fig:PN_models} depicts the phase noise spectra of the used models.}

\subsection{Phase noise compensation}\label{phase_noise_compensation}
A possible approach for the phase noise compensation is to average the phase error over a group of PTRS samples, then interpolate over the block for cancelling the phase error\cite{PN_tracking_Linear}.
The average phase error estimate over the $q^\text{th}$ group $\Bar{\Theta}_q$, is calculated as,
\begin{equation}
    \DM{\Bar{\Theta}_q = \text{arg} \Bigg(\frac{1}{N_P} \sum_{ m = 0}^{N_P-1} \frac{\mathbf{r}_q(m)\mathbf{p}^{\ast}_q(m)}{{|\mathbf{p}_q(m)|}^2}\Bigg)} 
\end{equation}
where $\mathbf{r}_q$ are the received phase tracking samples in the $q^\text{th}$ group and $\mathbf{p}_q$ are the corresponding transmitted PTRS samples. Then, the average values of all $Q$ groups can be interpolated to generate a phase error estimate $\mathbf{\Bar{\Theta}}$ of length $N_D + Q \times N_{P}$. The estimate is used to compensate for the PN by multiplying with the received signal as $\mathbf{r}(n)e^{-j\Bar{\mathbf{\Theta}}(n)}$.
Note that this method mainly removes the correlated PN contribution, which can readily be tracked. However, a residual amount of the PN remains uncompensated whose quantity depends on the occupied bandwidth and the uncorrelated PN level of the local oscillator. One could increase the PTRS density to better compensate for the PN errors, but since the Gaussian PN is uncorrelated, the pilot density would need to \DMMinorRev{be} increased significantly to fully compensate for the PN, which means useful resources for the data transmission are decreased by a considerable amount.

%% file: 4_Problemformulation.tex
In this section, we describe the optimization problem for waveform learning 
using the system model in Section \ref{3_SystemModel}. We consider the SC system in the Figure \ref{fig:ened_to_end_model}, \DM{in which} the constellation, transmit and receive filters are trainable. We will optimize the constellation and filters to ensure successful demapping under three demappers at the receiver: the conventional AWGN optimal demapper, a PN-aware demapper and then, we consider a trainable NN demapper $\mathbf{\mathbf{\Psi}}$, in place of the analytical demappers aiming to optimize the system further.

It has been shown that maximizing the bit-metric decoding rate can be achieved equivalently by minimizing the total binary cross entropy (BCE) loss \cite{FaycalPilotless}, which is commonly used in binary classification problems. The total binary cross entropy loss ${\mathcal{ L}}$ is expressed as,
\begin{align}
\begin{split}
\mathcal{ L}({\mathcal{ C}} )= {}& -\frac {1}{N_D} \sum _{n=0}^{N_D-1} \sum _{k=0}^{K-1}  b_{n,k} \mathop {\mathrm {log_{2}}} (\Gamma_{n,k} ) \\ 
&+ (1-b_{n,k}) \mathop {\mathrm {log_{2}}} (1-\Gamma_{n,k} ),  \\
\end{split}  
\end{align}
where $b_{n,k}$ and $\Gamma_{n,k}$ are the corresponding bits and the soft-estimates, 
when the trainable constellation $\mathcal{ C}$, trainable filters $\mathbf{g}_{tx}, \mathbf{g}_{rx}$  and the trainable demapper $\mathbf{\Psi}$ are used. Note that in case the AWGN optimal demapper and the PN demappers are used, $\mathbf{\Psi}$ is non-trainable. 
The BCE loss is estimated using multiple batches as usually done in the machine learning context.

We have demonstrated that by shaping the constellation we can achieve a reduction in the PAPR in our previous work\DM{\cite{ICCconstellation}}. One other key component that can be exploited to reduce PAPR, is the band-limiting pulse shaping filter at the transmitter and its corresponding receive filter in the receiver by trading-off stopband response.  However, regulatory bodies and standards define a limit on the maximum leakage power outside the bandwidth allocated to minimize the interference caused for the other bands. This has to be taken in to account when optimizing the waveform by ensuring the ACLR of the transmit waveform within a given bandwidth allocation is below a given limit. Therefore, our optimization problem is to learn the waveform in an end-to-end fashion by minimizing the BCE loss subject to a  PAPR limit and an ACLR limit for a given excess bandwidth for the filter, in the presence of phase noise. 
Formally this could be stated as,
\begin{subequations}\label{optimization_problem}
\begin{align}
&\underset{\mathcal{C}, \mathbf{g}_{tx}, \mathbf{g}_{rx}, \mathbf{\Psi}}{\text{minimize}}~\mathcal{ L}({\mathcal{ C}}, \mathbf{g}_{tx} , \mathbf{g}_{rx}, \mathbf{\Psi} ) \\
&\text{subject to}~\mathbb{E}_{c_i \in\mathcal{C}} \left[ c_i \right] = 0 \label{const_mean_condition}\\
&\hphantom{\text{subject to}~}\mathbb{E}_{c_i \in\mathcal{C}} \left[\left| c_i \right|^2\right] = 1 \label{const_energy_condition}\\
&\hphantom{\text{subject to}~} \sum_{g_i \in \mathbf{g}_{tx}} \left|g_i \right|^2  = 1  \label{filter_energy_condition}\\
&\hphantom{\text{subject to}~} \text{PAPR}(\mathbf{g}_{tx}, \mathcal{C}) \leq \epsilon_P \label{papr_constraint}\\
&\hphantom{\text{subject to}~} \text{ACLR}_{\beta}(\mathbf{g}_{tx}) \leq \epsilon_A \label{aclr_constraint} 
\end{align}
\end{subequations}
The condition in \eqref{const_mean_condition} ensures a centred constellation on the IQ-plane, while \eqref{const_energy_condition} ensures that the learned constellation is normalized in energy. A normalized transmit filter is established by \eqref{filter_energy_condition}. The PAPR is a function of both constellation points and the transmit filter, which is enforced in the condition in \eqref{papr_constraint} to be below $\epsilon_P$. The ACLR constraint stated in \eqref{aclr_constraint} ensures that the out-of-band leakage of the transmit filter outside a $1+\beta$ bandwidth is below a threshold $\epsilon_A$, where $\beta$ is the excess bandwidth.
 
\subsection{Trainable constellation}
A trainable constellation, which uses 2D signalling points on the IQ-plane can be modelled by a set of ${2^K}$  trainable complex weights $\tilde {\mathcal{C}}$, where $K$ is the number of bits.
Such a constellation $\mathcal{C}$ is described by \cite{FaycalPilotless,WaveformFaycal},

\begin{equation} 
    \mathcal{C}= \frac{\tilde {\mathcal{C}} - \frac{1}{2^{K}}\sum_{c_i \in \tilde {\mathcal{C}}}^{}c_i}{\sqrt{\frac{1}{2^{K}}|\tilde {\mathcal{C}} - \frac{1}{2^{K}}\sum_{c_i \in \tilde {\mathcal{C}}}^{}c_i|^2}} 
\label{eq_trainable_constellation}
\end{equation}

The bit labelling is set by the binary value of the index of each element in $\mathcal{C}$. The bit to symbol mapping is done as in a conventional constellation. Therefore, no additional complexity is introduced when it is used in the transmitter compared to the traditional mapping. However, it allows us to optimize the constellation points and the corresponding bit-labels. During training, the constellation points will traverse on the IQ-plane and rearrange themselves optimally to meet the optimization criteria.

\subsection{Trainable filters}
A trainable FIR filter impulse response with a set of $L$ real-valued trainable weights $\Tilde{\mathbf{g}}$ can be defined as, 
\begin{equation}
    \mathbf{g} = \frac{\Tilde{\mathbf{g}}}{\sqrt{\sum_{g_i\in \Tilde{\mathbf{g}}} {|g_i|}^2}}
\end{equation}
The span of the filter is $L = S\times M+1$, where $S$ is the filter span in symbols and $M$ is the oversampling factor. 
The filter impulse response is trained with energy normalization, which fulfills the condition in \eqref{filter_energy_condition}. Moreover, the frequency response of the filter is symmetric since a real-valued filter is desired. We consider two separate filters for the transmit and receive sides as $\mathbf{g}_{tx}$ and $ \mathbf{g}_{rx}$. 
\subsection{Demappers}

\subsubsection{AWGN Optimal Demapper (AOD)}
In a conventional system, the demapper is a fixed function, where the soft estimates are the LLRs calculated by \DM{assuming} that the signals are corrupted by AWGN noise, which we refer as the AWGN optimal demapper. 
\DM{These estimates are clearly suboptimal, since they are calculated neglecting the residual Gaussian PN existing in most of the practical systems operating in high frequency range. The higher the residual PN level remaining in the system, the higher the error level of the estimates - as also shown in \cite{Spiral} for the standard-QAM constellation case. Nevertheless, we optimize the waveform under the AOD, intending to identify constellation and filter pairs, which show the most resilience to such a mismatch. We aim to assess the best possible performance of a conventional system design when scaled up to apply in higher frequency range (sub-THz and THz) in a straightforward manner. This system performance can be used as a reference to compare with those of other approaches. }
For the AOD, the soft estimate or the LLR for each bit is calculated\DM{\footnote{Tensorflow single-precision implementation of LogSumExp \texttt{tf.math.reduce\_logsumexp} was used for computation.}} from \eqref{eq_LLR_main} as,
\begin{equation}
    \Gamma_{n,k} = \log \left( \frac{\sum_{c_i\in \mathcal{C}(1)} \exp \left(-\frac{|{\mathbf{r}}(n)-c_i|^2}{\sigma^2_n}\right)}{\sum_{c_i\in \mathcal{C}(0)} \exp \left(-\frac{|{\mathbf{r}}(n)-c_i|^2}{\sigma^2_n}\right)} \right),
\label{eq_AWGN_demapper}
\end{equation}
\DM{where $\sigma^2_n$ is the complex Gaussian noise variance.}
 The AWGN noise variance $\sigma^2_n$ for a certain $\frac{E_b}{N_0}$ is calculated as,
\begin{equation}
    \sigma^2_n = \Big[ \Big(\frac{E_b}{N_0}\Big) r M \Big(\frac{N-QN_P}{N+N_{CP}}\Big)\Big]^{-1},
\end{equation}
where $r$ is the code rate. 

\subsubsection{Phase Noise Demapper (PND)}

A better alternative to the conventional estimate described above is to explicitly consider the residual Gaussian PN in the LLR estimate. \DMMinorRev{The exact derivation of the likelihood function under Gaussian PN is challenging due to its dependence on the posteriori phase probability density function (PDF), for which obtaining a closed-form expression is analytically intractable \cite{KrishnanSoftMetrics}.  As mentioned earlier, a pragmatic solution to this is to use approximate maximum likelihood detectors derived in \cite{KrishnanPNOptimization} and \cite{KrishnanSoftMetrics}. The approximations utilize a high instantaneous SNR assumption meaning, $|r(n)|\gg\Re\{w(n)\}$, or low instantaneous PN assumption allowing the use of the  first-order Taylor approximation, $e^{j\Theta(n)} \approx 1 + j \Theta(n)$. Then, the likelihood functions are derived with simple formulations, which can be utilized as PN-aware demapping rules.} 
The log likelihood function under low PN (PND-LPN) is  \cite{KrishnanPNOptimization}, 
\begin{align}\label{lowPN_LLF}
    \log(p(r|c)) &= -\frac{1}{\sigma^2_n}(\Re\{re^{-j\arg(c)}\}-\vert c\vert)^{2} \notag\\
    &-\frac{(\Im\{re^{-j\arg(c)}\})^{2}}{\sigma_{\rm p}^{2}\vert c\vert ^{2}+\sigma^2_n}-\log(\sigma_{\rm p}^{2}\vert c \vert^{2}+\sigma^2_n)+\mathrm{c}.\mathrm{c},
\end{align}
while for the high SNR (PND-HSNR) approximation is,
\begin{align}\label{highSNR_LLF}
    \log(p(r|c)) &= -{(\vert r \vert-\vert c\vert)^{2}\over \sigma^2_n} \notag\\
    &-{\big(\arg(r)-\arg(c)\big)^{2}\over \sigma_{\rm p}^{2}+\frac{\sigma^2_n}{\vert c\vert^{2}}}-\log\left(\sigma_{\rm p}^{2}\vert c\vert^{2}+\sigma^2_n\right)+\mathrm{c}.\mathrm{c},
\end{align}
where $\sigma_{\rm p}^{2}$ is the residual PN variance. \DM{Then, the constellation and filters are tailored to optimize the performance of the system designed applying these demappers using the log likelihood function in \eqref{eq_LLR_main}. \DMMinorRev{Note that, the demapping rules require knowledge on both variances of the AWGN and residual PN.}
Since we emulate a realistic PN model, for the measurement of residual PN variance, we utilize another set of distributed pilots, additionally to the PTRS described earlier, which is appended to the PTRS blocks with each block having $N_R$ samples per block. This is taken appropriately into account when calculating the ${E_b}/{N_0}$ and spectral efficiencies in evaluations. In the receiver, after performing PTRS based Wiener PN compensation as described in subsection \ref{phase_noise_compensation}, the received pilots for residual PN are extracted. When the transmitted residual PN pilots are $\mathbf{u}$, and the received residual PN pilots are $\mathbf{v}$, the residual PN variances can be estimated with maximum likelihood estimators derived based on \eqref{lowPN_LLF} for low PN as,}

\begin{align} 
\hat{\sigma }_{n_{LPN}}^2 &= \frac{1}{{QN_{R}}} \sum _{l=1}^{QN_{R}} (\Re\{\mathbf{v}(l)e^{-j\arg(\mathbf{u}(l))}\}-\sqrt{E_s})^2, \notag\\ 
\hat{\sigma }_{p_{LPN}} ^2 &= \frac{1}{{QN_{R}}} \sum _{l=1}^{QN_{R}} \big(\Im\{\mathbf{v}(l)e^{-j\arg(\mathbf{u}(l))}\}\big)^2 - \frac{\hat{\sigma }_{n_{LPN}}^2}{E_s} 
\end{align}
and based on \eqref{highSNR_LLF} for high SNR as,  
\begin{align} 
\hat{\sigma }_{n_{HSNR}}^2 &= \frac{1}{{QN_{R}}} \sum _{l=1}^{QN_{R}} (\vert \mathbf{v}(l) \vert-\sqrt{E_s})^2, \notag\\ 
\hat{\sigma }_{p_{HSNR}} ^2 &= \frac{1}{{QN_{R}}} \sum _{l=1}^{QN_{R}} \big(\arg(\mathbf{v}(l))-\arg(\mathbf{u}(l))\big)^2 - \frac{\hat{\sigma }_{n_{HSNR}}^2}{E_s}, 
\end{align}
where $E_s$ is the average signal power.


\subsubsection{Neural network demapper (NND)}
\begin{figure}[h]
    \centering
    \includegraphics[width=\columnwidth]{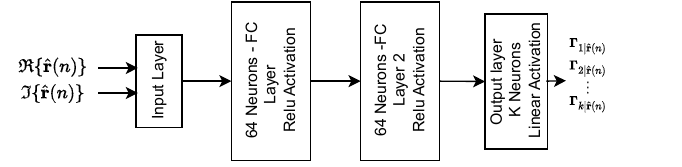}
    \caption{Neural Demapper}
    \label{fig:NNDemapper}
\end{figure}

To further explore trainability of the system we consider a simple neural network as the demapper. This is also motivated by the fact that optimal maximum likelihood detector is difficult to derive in exact form and the PNDs are approximations\cite{KrishnanPNOptimization}. The neural network demapper considered is a simple fully-connected neural network with a linear layer at the output to generate soft-estimates as shown in Figure \ref{fig:NNDemapper}, which will be trained in the end-to-end model. 
This allows the demapper to adapt to the noise conditions, jointly with the constellation and the filters.  Moreover, it has been shown that such a simple neural network based demapper can be successfully used to mimic \DMMinorRev{the} traditional demappers with low complexity \cite{JakobLLR} as well. The input to the demapper consists of real and imaginary parts of the received symbol, while the outputs are the $K$ soft-estimates for each bit considering the BMD scheme.

\subsection{Constraints}
In this sub-section, we focus on the constraints of the problem in \eqref{optimization_problem}. The conditions \eqref{const_mean_condition} and \eqref{const_energy_condition}  are implicitly enforced by the model for the constellation. The power characteristics of the waveform depends on the distribution of the constellation points as well as the transmit filter impulse response. Note that phase noise does not change the power of the transmit signal since it distorts only the phase. The PAPR is best described in a probabilistic sense \cite{PAPR} as,
\begin{equation}
    \DM{\text {PAPR} = {\text{min}} \bigg\{ \nu \: \big| \:\Pr \left ({\frac {{|\mathbf{\hat{s}}(n)|}^2}{ \mathbb{E}[{|\mathbf{\hat{s}}(n)|}^2]} > \nu }\right) \leq \delta_p \bigg\}}
\end{equation}
The PAPR in its conventional meaning can be obtained by setting $\delta_{P} = 0$.
In the PAPR constraint in \eqref{papr_constraint}, we aim to have $\text{PAPR}(\mathcal{C}) \leq \epsilon_P$, which means we ensure that the threshold $\nu $ is our desired PAPR $\epsilon_P$, which can be equivalently expressed as, 
\begin{equation}
    \mathbb{E}\bigg( \text{max}\Big(\frac {{|\mathbf{\hat{s}}(n)|}^2}{ \mathbb{E}[{|\mathbf{\hat{s}}(n)|}^2]}-\epsilon_P, 0 \Big)\bigg)  = 0,
\end{equation}
where the expectation is over the transmit signal $\mathbf{\hat{s}}(n)$. Therefore the PAPR constraint, $\Phi_P$ can be computed by Monte Carlo sampling of the transmit signal $\mathbf{\hat{s}}(n)$ during training as,
\begin{align}\label{PAPR_const_MC}
    \Phi_P (\mathcal{C}, \mathbf{g}_{tx}, \epsilon_P) &\approx \frac{1}{V}\sum_{v=1}^{V} \text{max}\Big(\frac {{|\mathbf{\hat{s}}_v(n)|}^2}{ \mathbb{E}[{|\mathbf{\hat{s}}(n)|}^2]}-\epsilon_P, 0 \Big) \notag\\
    &= 0,
\end{align}
where $V$ is the number of power samples.

The band-limiting transmit filter $\mathbf{g}_{tx}$ must ensure that the out-of-band emission does not exceed a certain level. The adjacent channel leakage ratio (ACLR) is the ratio of the stopband energy ($\xi_S$) to the in-band energy ($\xi_I$) of the transmit filter. First, we define the in-band frequencies associated with the transmit filter as $\big[-\frac{1+\beta}{2}, \frac{1+\beta}{2}\big]$, where $\beta$ is the excess bandwidth outside the Nyquist bandwidth similar to RRC filters. The filter is designed for an oversampling rate of M, which means the stop-band energy outside the $1+\beta$ bandwidth can be computed by \cite{NyquistM},
\begin{equation}\label{eq_stopband}
    \mathcal{\xi}_S = \mathbf{g}_{tx}^\mathrm{T} \mathbf{\Phi} \mathbf{g}_{tx},
\end{equation}
where $\mathbf{\Phi}= [\Phi_{L \times L}]$ with,
\begin{equation}
    \phi_{nm}=\begin{cases} \begin{aligned} 
    & 1-\frac{1+\beta}{M}, \quad n=m,\\
    & -\frac{1+\beta}{M} \text{sinc}\Big(\frac{1+\beta}{M}(n-m)\Big), \quad n \ne m.
\end{aligned} \end{cases}    
\end{equation}
Since the total energy, $\xi_I + \xi_S = 1$ the $\text{ACLR}_\beta$, which means the leakage ratio outside the $1+\beta$ bandwidth can be computed with,
\begin{align}
    \text{ACLR}_\beta &= \frac{\xi_S}{\xi_I} =  \frac{\xi_S}{1-\xi_S}
\end{align}
Then the ACLR constraint in \eqref{aclr_constraint}, for a certain $1+ \beta$ can be defined as,
\begin{equation}
    \Phi_A(\mathbf{g}_{tx}, \beta, \epsilon_A) = \text{ACLR}_{\beta} - \epsilon_A \leq 0                                      
\end{equation}
Note that the conventional definition of ACLR considers the normalized Nyquist bandwidth $[-\frac{1}{2}, \frac{1}{2}]$, which can be evaluated by substituting $\beta=0$.

With these expressions, the optimization problem \eqref{optimization_problem} can be rewritten with \eqref{const_energy_condition} and \eqref{filter_energy_condition} conditions being implicit as, 
\begin{subequations}\label{optimization_problem_reform}
\begin{align}
&\underset{\mathcal{C}, \mathbf{g}_{tx}, \mathbf{g}_{rx}, \mathbf{\mathbf{\mathbf{\Psi}}}}{\text{minimize}}~\mathcal{ L}({\mathcal{ C}}, \mathbf{g}_{tx} , \mathbf{g}_{rx}, \mathbf{\mathbf{\Psi}} ) \\
&\hphantom{\text{subject to}~} \Phi_P (\mathcal{C},\mathbf{g}_{tx}, \epsilon_P) = 0 \label{papr_constraint_reform} \\
&\hphantom{\text{subject to}~}  \Phi_A(\mathbf{g}_{tx}, \beta, \epsilon_A) \leq 0 \label{aclr_constraint_reform} 
\end{align}
\end{subequations}

%% file: 5_OptimizationMethodology.tex
\label{Section_5_Optimization}
The problem of constellation shaping for maximizing the information rate itself is a non-convex problem \cite{noncovex}. Therefore, the problem we are trying to solve, which considers both constellation and filters is a non-convex constrained optimization problem. Since the goal is to optimize the system in an end-to-end fashion, we employ the augmented Lagrangian method for solving the problem, which allows reformulating the BCE loss and the constraints in the problem \eqref{optimization_problem_reform} into a differentiable loss function. With the construction of the augmented Lagrangian, the constrained optimization problem can be solved with a series of unconstrained problems using stochastic gradient descent (SGD) by computing the gradients and back propagating through the system with respect to the trainable parameters. \DM{This optimization method has been successfully used 
in \cite{WaveformFaycal} and \cite{MathieuConference} for similar problems. }

The inequality constraint in \eqref{aclr_constraint_reform} is first converted to an equality constraint by introducing a slack variable $z$, which translates the problem \eqref{optimization_problem_reform} to an equality constraint problem as,
\begin{subequations}\label{optimization_problem_reform_equality}
\begin{align}
&\underset{\mathcal{C}, \mathbf{g}_{tx}, \mathbf{g}_{rx}, \mathbf{\mathbf{\Psi}}}{\text{minimize}}~\mathcal{ L}({\mathcal{ C}}, \mathbf{g}_{tx} , \mathbf{g}_{rx}, \mathbf{\Psi} ) \\
&\hphantom{\text{subject to}~} \Phi_P ( \mathcal{C}, \mathbf{g}_{tx},\epsilon_P) = 0  \\
&\hphantom{\text{subject to}~}  \Phi_A(\mathbf{g}_{tx}, \beta, \epsilon_A) + z =  0 
\end{align}
\end{subequations}
\subsection{Augmented Lagrangian}
The augmented Lagrangian can be then written for $\lambda > 0$ as,
\begin{multline}\label{augmented_lagrangian}
\mathcal{ L}_{aug}({\mathcal{ C}}, \mathbf{g}_{tx} , \mathbf{g}_{rx}, \mathbf{\Psi}, z, \mu_P, \mu_a, \lambda )=\mathcal{ L}({\mathcal{ C}}, \mathbf{g}_{tx} , \mathbf{g}_{rx}, \mathbf{\Psi} ) \\
+ \mu_P\Phi_P ( \mathcal{C}, \mathbf{g}_{tx}, \epsilon_P)  + \mu_A \Phi_A(\mathbf{g}_{tx}, \beta , \epsilon_A) \\
 + \frac{\lambda}{2}\Big[ |\Phi_P( \mathcal{C}, \mathbf{g}_{tx}, \epsilon_P)|^2 + |\Phi_A(\mathbf{g}_{tx}, \beta, \epsilon_A) + z|^2 \Big],
\end{multline}
where $\mu_P$ and $\mu_A$ are the Lagrangian multipliers associated with the PAPR and ACLR constraints and $\lambda$ is the penalty parameter. The minimization of the augmented Lagrangian with respect to the slack variable $z$ is quadratic in $z$, for fixed $\mathcal{ C}, \mathbf{g}_{tx} , \mathbf{g}_{rx}, \mathbf{\Psi}$, which allows the minimum to be calculated in-place \cite[Chapter~3]{augemented_book}. This converts the augmented Lagrangian to,
\begin{multline}\label{augmented_lagrangian_reform}
\mathcal{ L}_{aug}({\mathcal{ C}}, \mathbf{g}_{tx} , \mathbf{g}_{rx}, \mathbf{\Psi}, \mu_P, \mu_A, \lambda ) =  \mathcal{ L}({\mathcal{ C}}, \mathbf{g}_{tx} , \mathbf{g}_{rx}, \mathbf{\Psi} ) \\
+ \mu_P\Phi_P ( \mathcal{C}, \mathbf{g}_{tx}, \epsilon_P)  
 + \frac{\lambda}{2} |\Phi_P ( \mathcal{C}, \mathbf{g}_{tx}, \epsilon_P)|^2\\  + \frac{1}{2\lambda} \Big[max \{ 0, \mu_A+\lambda\Phi_A(\mathbf{g}_{tx}, \beta, \epsilon_A)]^2 -\mu_A^2\} \Big]
\end{multline}

\subsection{System training}
Since we apply tools from AI/ML community, we use the term ``training'' and optimizing throughout this paper interchangeably.
The optimization procedure involves iterative minimization of the augmented Lagrangian in \eqref{augmented_lagrangian_reform} by SGD, which we outline in the following Algorithm  \ref{train_alg}. The initial $\textit{System Parameters}$ include the carrier frequency $f_c$ and bandwidth, number of bits $K$, data block length $N_D$, PTRS parameters $Q$ and $N_P$, CP length $N_{CP}$, oversampling factor $M$, filter lengths $L$, excess bandwidth ($\beta$), PN model parameters on transmitter and receiver sides and the NND hyper parameters if it is used. 
The end-to-end model comprises all the functionality inside the dashed box indicated in Figure \ref{fig:ened_to_end_model}. During the SGD steps, randomly generated data bits are carried through the end-to-end model to generate bitwise soft-estimates, which are used to calculate the BCE loss. Then we have a fully differentiable loss function, which is used to compute gradients w.r.t. each of the trainable parameters and back propagated through the system to update the trainable parameters.

\begin{algorithm}[ht]
\caption{Training algorithm}\label{train_alg}
\begin{algorithmic}
\STATE $\textbf{Set} \;\; \textit{\{System Parameters\}}$
\STATE $\textbf{Initialize} \;\; {\mathcal{ C}}, \mathbf{g}_{tx} , \mathbf{g}_{rx}, \mathbf{\Psi}, \mu_P^{[0]}, \mu_A^{[0]}, \lambda^{[0]} $
\STATE $\textbf{for} \;\; i = 0, \dots \textbf{do}$
\STATE \hspace{0.5cm}$\triangleright \: \text{Perform SGD}$
\STATE \hspace{0.5cm}$\textbf{for} \;\; j = 0, \dots \textbf{do}$
\STATE \hspace{0.5cm}\hspace{0.5cm}$ \textbf{forward pass}: \text{ from} \;\mathbf{B}^{[j]} \; \text{to} \; \mathbf{\Gamma}^{[j]} $
\STATE \hspace{0.5cm}\hspace{0.5cm}$ \textbf{compute loss}: \mathcal{ L}^{[j]}({\mathcal{ C}}, \mathbf{g}_{tx} , \mathbf{g}_{rx}, \mathbf{\Psi} ) $ 
\STATE \hspace{0.5cm}\hspace{0.5cm}$ \textbf{compute}: \Phi^{[j]}_P ( \mathcal{C}, \mathbf{g}_{tx}, \epsilon_P), \Phi^{[j]}_A (\mathbf{g}_{tx}, \beta, \epsilon_A)  $
\STATE \hspace{0.5cm}\hspace{0.5cm}$ \textbf{compute}: \mathcal{ L}_{aug}^{[j]}({\mathcal{ C}}, \mathbf{g}_{tx} , \mathbf{g}_{rx}, \mathbf{\Psi}, \mu_P^{[i]}, \mu_A^{[i]}, \lambda^{[i]} )$
\STATE \hspace{0.5cm}\hspace{0.5cm}$ \textbf{backward pass}:  \partial \mathcal{ L}_{aug}^{[j]} $
\STATE \hspace{0.5cm}$\textbf{end for}$
\STATE \hspace{0.5cm}$\triangleright \: \text{Update Lagrangian Multipliers}$
\STATE \hspace{0.5cm}$ \textbf{Recompute}: \Phi^{[i]}_P ( \mathcal{C}, \mathbf{g}_{tx}, \epsilon_P), \text{ACLR}^{[i]}_c(\mathbf{g}_{tx}, \beta, \epsilon_A)  $
\STATE \hspace{0.5cm}$\mu_P^{[i+1]}\leftarrow \mu_P^{[i]} + \lambda^{[i]} \Phi^{[i]}_P ( \mathcal{C}, \mathbf{g}_{tx}, \epsilon_P) $
\STATE \hspace{0.5cm}$\mu_A^{[i+1]}\leftarrow max \{ 0, \mu_A^{[i]}+\lambda^{[i]}\Phi_A^{[i]}(\mathbf{g}_{tx}, \beta, \epsilon_A)\}  $
\STATE \hspace{0.5cm}$\triangleright \: \text{Update penalty parameter}$
\STATE \hspace{0.5cm}$\lambda^{[i+1]} \leftarrow \tau\lambda^{[i]} \: \text{where} \:\tau>1$
\STATE $\textbf{end for}$
\end{algorithmic}
\label{alg1}
\end{algorithm}

%% file: 6_EvaluationSetup.tex
The implementation was done in the widely-used machine learning framework Tensorflow and some components from the open-source library for physical layer research – Sionna \cite{Sionna}. We use the automatic differentiation feature and the optimizers to implement the SGD step. Table \ref{sim_table} summarizes the simulation parameters used for the evaluation.

\begin{table}[ht] \caption{Simulation parameters}\label{sim_table}
\begin{tabular}{|l|l|}
\hline
\textbf{Carrier Frequency}                                                        & 120, 220 GHz                                                                    \\ \hline
\textbf{Bandwidth}                                                                 & 3.93 GHz                                                                   \\ \hline
\textbf{Block size$(N)$}                                                           & 4096                                                                          \\ \hline
\textbf{Modulation order $(K)$}                                                    & 6                                                                                  \\ \hline
\textbf{Oversampling factor  $(M)$}                                                 & 4                                                                                     \\ \hline
\textbf{RRC filter span  $(L)$}                                                     & 32 symbols                                                                            \\ \hline
\textbf{RRC roll-off   $(\beta)$}                                                     & 0.3                                                                           \\ \hline
\textbf{CP Ratio   $(N_{CP}/(N+N_{CP}))$}                                           & 7.03125 \%                                                                \\ \hline
\textbf{No of. PTRS blocks  $(Q)$}                                                          & 32                                                                                 \\ \hline
\textbf{\begin{tabular}[c]{@{}l@{}}PTRS symbols per\\ PTRS block $(N_P)$\end{tabular}}& 4                                                                                     \\ \hline
\textbf{\begin{tabular}[c]{@{}l@{}}RPN pilots per\\ PTRS block $(N_R)$\end{tabular}}& 1 - for 120 GHz, 4 - for 220 GHz                                                           \\ \hline
\textbf{PTRS and RPN symbols}                                                               & Zadoff-Chu sequences                                                                  \\ \hline\hline
\textbf{PAPR Targets $(\epsilon_P)$}                                                & 5.5 dB, 6.0 dB, 6.5 dB                                         \\   \hline
\textbf{ACLR Targets $(\epsilon_A)$}                                                                & -45 dB, -55 dB for $1+ \beta$                            \\ \hline
\textbf{Excess bandwidth targets ($\beta$ )}                                       & 0.25, 0.3                                                      \\ \hline
\textbf{No. of power samples (V)}                                                   & $4\times 10^5 $                                                                                   \\ \hline
\textbf{Batch size}                                                                 & 10                                                                                    \\ \hline
\textbf{Optimizer}                                                                  & Adam                                                                                  \\ \hline
\textbf{Learning rate}                                                              & 1e-3                                                                                  \\ \hline
\textbf{Training EbN0 range}                                                        & {[}6-18{]} dB                                         \\ \hline
\textbf{EbN0 sampling}                                                              & Uniform                                                                               \\ \hline \hline
\textbf{Evaluation – Channel Code}                                                  & 5G NR LDPC\DM{\tablefootnote{Implementation in Sionna\cite{Sionna}.}}        \\ \hline
\textbf{Code rate}                                                                  & 3/4                                                                                   \\ \hline
\textbf{BP iterations}                                                              & 50                                                                                    \\ \hline
\end{tabular}
\end{table}

\DMMinorRev{The constellation was initialized with a 8+16+20+20 amplitude and phase shift keying (APSK) constellation with quasi-Gray-labelling, which is chosen for its deployment in the digital video broadcasting standard by the European telecommunications standards institute (ETSI) (DVB-S2) \cite{standard2020digital}}, while the filters are initialized by an RRC with the respective $\beta$ value.  We train the end-to-end system on a range of $E_b/N_0$ values with uniform sampling at each iteration such that the learned waveform is robust in the  applicable SNR range for a chosen modulation order. The end-to-end system is trained as an uncoded ($r=1$) system and for evaluation, an outer channel code with 5G NR \DMMinorRev{low density parity check (LDPC)} was used for computing the BLER of the system and the effective bandwidth ($BW_{eff} = \text{normalized occupied bandwidth (OBW)} \times \text{Bandwidth} $), where 99.9\% of the power is contained for the trained system. Then the spectral efficiency ($SE$) is calculated as,

\begin{equation}
    SE= \frac{(1-BLER)*R}{BW_{eff}}
\end{equation}
and $R$ is the data rate given by,
\begin{equation}
    R = r M \Big(\frac{N-QN_P}{N+N_{CP}}\Big)\frac{1}{T_s},
\end{equation}
where $T_s$ is the symbol duration. Note that all the performance evaluations are done with a fixed code rate of 3/4.

\begin{table}[ht]\caption{Tx PN model - TI LMX2595}\label{TxPN_model}
\centering
\begin{tabular}{|c|cccc|}
\hline
PSD0 & \multicolumn{4}{c|}{$6.3096 \times 10^{-8}$ (-72 dB)}                                                    \\ \hline
n,m  & \multicolumn{1}{c|}{$f_{z,n}$}  & \multicolumn{1}{c|}{$\alpha_{z,n}$} & \multicolumn{1}{c|}{$f_{p,n}$}   & $\alpha_{p,n}$ \\ \hline
1    & \multicolumn{1}{c|}{$3\times10^{-6}$}    & \multicolumn{1}{c|}{1.4}  & \multicolumn{1}{c|}{10}   & 1    \\ \hline
2    & \multicolumn{1}{c|}{$1.75\times 10 ^7$} & \multicolumn{1}{c|}{2.55} & \multicolumn{1}{c|}{$3.00 \times 10^5$}  & 2.95 \\ \hline
\end{tabular}
\end{table}

\begin{table}[ht] \caption{Rx PN model - 3GPP TR38.803 v14.2.0  UE model 1} \label{Rx_PN_model}
\centering
\begin{tabular}{|c|cccc|}
\hline
       & \multicolumn{4}{c|}{ Loop BW = 187 kHz}                                                   \\ \hline
       & \multicolumn{1}{c|}{Ref clk} & \multicolumn{1}{c|}{PLL}      & \multicolumn{1}{c|}{$VCO_{V2}$}    & $VCO_{V3}$ \\ \hline
FOM    & \multicolumn{1}{c|}{-215}    & \multicolumn{1}{c|}{-240}     & \multicolumn{1}{c|}{-175}      & -130   \\ \hline
fz     & \multicolumn{1}{c|}{Inf}     & \multicolumn{1}{c|}{$1.00\times 10 ^4$} & \multicolumn{1}{c|}{$50.30 \times 10^6$} & Inf    \\ \hline
P (mW) & \multicolumn{1}{c|}{10}      & \multicolumn{1}{c|}{20}       & \multicolumn{2}{c|}{20}                 \\ \hline
k      & \multicolumn{1}{c|}{2}       & \multicolumn{1}{c|}{1}        & \multicolumn{1}{c|}{2}         & 3      \\ \hline
\end{tabular}
\end{table}

\DMMinorRev{
We consider the measured PN PSD of a Texas Instruments LMX2595 at $f_c = 20 \text{ GHz}$ (Figure 14 in \cite{ti_lmx2595_datasheet}), approximated by polynomials with Equation \eqref{PN_3GPP_model_10} for the transmitter and the used parameters are given in Table \ref{TxPN_model}. The receiver PN is modelled using the parameters in Table 6.1.11.2-1 in 3GPP TR38.803 v14.2.0 \cite{3gpp.38.803}, which is given in Table \ref{Rx_PN_model}. }

%% file: 7_Results.tex
In this section, we present numerical results pertaining to the waveform learning problem. Initially, we showcase the learned constellations and filters. Subsequently, we delve into their PAPR, BLER, and SE performance, comparing them with baseline solutions employing conventional gray-labeled QAM and a 8+16+20+20 APSK constellation.  \DMMinorRev{The latter from ETSI DVB-S2 \cite{standard2020digital} was identified as the best-performing APSK scheme in terms of both BLER and PAPR for our settings.} Furthermore, the performance of pulse shaping filters is assessed in comparison to a pair of RRC filters with a roll-off/excess bandwidth of 0.3. BLER and SE metrics are evaluated concerning the mentioned baseline waveforms, utilizing both AOD and the PND at the receiver.

\subsection{PAPR characteristics}
\begin{figure}
    \centering
    \includegraphics[width=0.85\columnwidth]{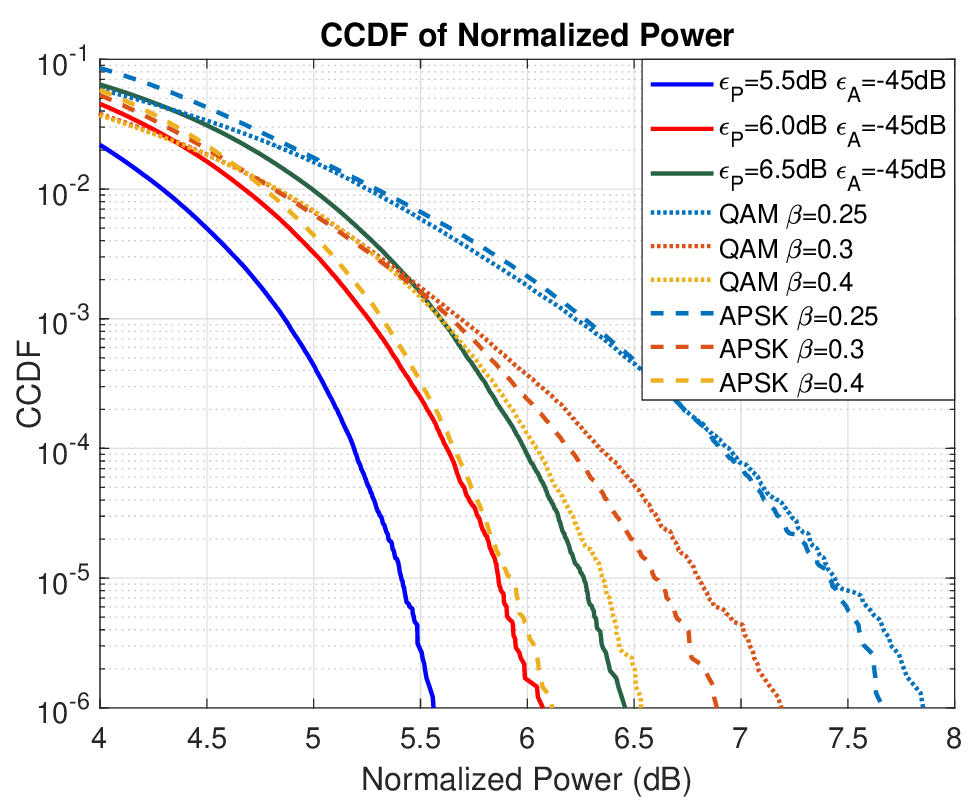}
    \caption{CCDF of the normalized power for waveforms learned under different $\epsilon_P$ targets, using NND at 120 GHz with $\epsilon_A = -45\text{ dB}$ and $\beta = 0.3$. The baseline waveforms shown are using QAM and APSK with RRC $\beta=0.25, 0.3,  0.4$}
    \label{fig:CCDF}
\end{figure}

To investigate the power characteristics, the complementary cumulative distribution function (CCDF) of the normalized power samples from the learned waveforms are presented in Figure \ref{fig:CCDF} for different $\epsilon_P$ targets. The CCDF for each curve is plotted with approximately $8\times10^6$ samples, while during training, we utilize about $V= 4\times10^5$ samples per each iteration in the calculation of \eqref{PAPR_const_MC}. The depicted curves are for the case of waveforms learned with the NND at 120 GHz under $\epsilon_A = -45\text{ dB}$ with a $\beta = 0.3$. For comparison, the power characteristics of the waveforms with APSK and QAM with $\beta = 0.3~\text{and}~0.4$ values are also included, along with the case where the \DMMinorRev{learned} waveform without a PAPR target is presented. 
It is evident that the learned waveforms adhere to the $\epsilon_P$ targets. At the CCDF value of $10^{-5}$, the normalized power samples are well below $\epsilon_P$ target.  With a $\epsilon_P = 5.5$ dB, at the CCDF value of $10^{-5}$, we achieve a PAPR reduction of 1.2 dB compared to the baseline of APSK and RRC with $\beta = 0.3$. Notably, lower PAPR of $5.9 \text{ dB}$ for an RRC is obtained at the cost of more excess bandwidth with $\beta = 0.4$. However, with the learned waveform, we achieve an even lower PAPR utilizing a 0.3 excess bandwidth while maintaining competitive BLER and SE performance as we will demonstrate in the  subsection \ref{BLERandSE}.

\subsection{Learned Constellations}
\begin{figure*}
    \centering
    \includegraphics[width=\textwidth]{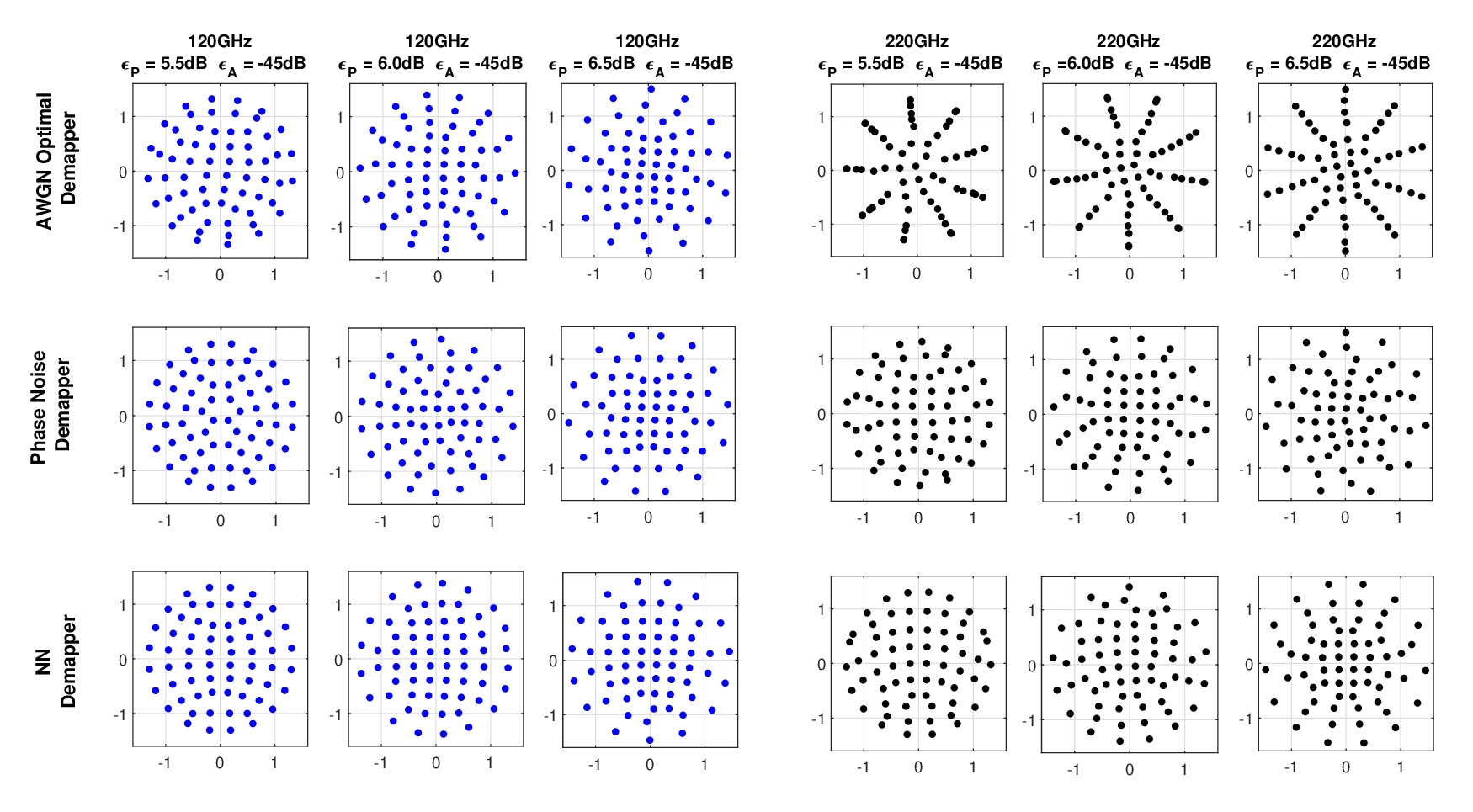}
    \caption{Learned constellations at 120 GHz and 220 GHz with $\beta =0.3$ $\epsilon_A = -45 \text{ dB}$ for different $\epsilon_P$ targets under different demappers}
    \label{fig:constellations}
\end{figure*}
Figure \ref{fig:constellations} illustrates the learned constellations for PAPR targets at 120 GHz (first three columns) and 220 GHz (last three columns), maintaining an ACLR$_\beta$ target of $\epsilon_A = -45 \text{ dB}$ and an excess bandwidth of $\beta = 0.3$. The rows correspond to the three demapping methods: AOD, PND, and NND employed at the receiver. Notably, all learned constellation point distributions exhibit rounded shapes reminiscent of APSK constellations, contributing to lower PAPR compared to the traditional QAM. Focusing on the 120 GHz constellations initially, under a strict PAPR constraint of $\epsilon_P = 5.5\text{ dB}$, a compact arrangement is observed at the outer regions restricting the maximum amplitude. As PAPR targets are relaxed, increased maximum amplitudes are observed. For a certain PAPR target across demapping methods, different constellations are observed, which reflects the impact of the demapping rule in the LLR calculation. Symmetry is not always observed, but a bilateral symmetry at least in I or Q axis can be seen often, specially in NND constellations. 

Moving to the 220 GHz cases, where the PN becomes stronger, a clear distinction between the demapping methods can be observed. In the AOD case, the utilization of Euclidean distance as the metric results in points arranging themselves with increased separation in the outer regions, creating a distinct branching effect. This arrangement leads to points on the branches being closer to each other, \DM{especially} when the PAPR target is strict. In contrast, with PND and NND, the point distributions exhibit a more relaxed pattern. Again, some bilateral symmetry is observed along I or Q axis. We initialized the bit-labelling from the APSK scheme, which has a quasi gray mapping. We observed that quasi-gray mapping is still preserved for the learned labelling although the constellation points move significantly in the IQ-plane compared to the initialization. 

\subsection{Learned filters}
\begin{figure}
    \centering
    \includegraphics[width=\columnwidth]{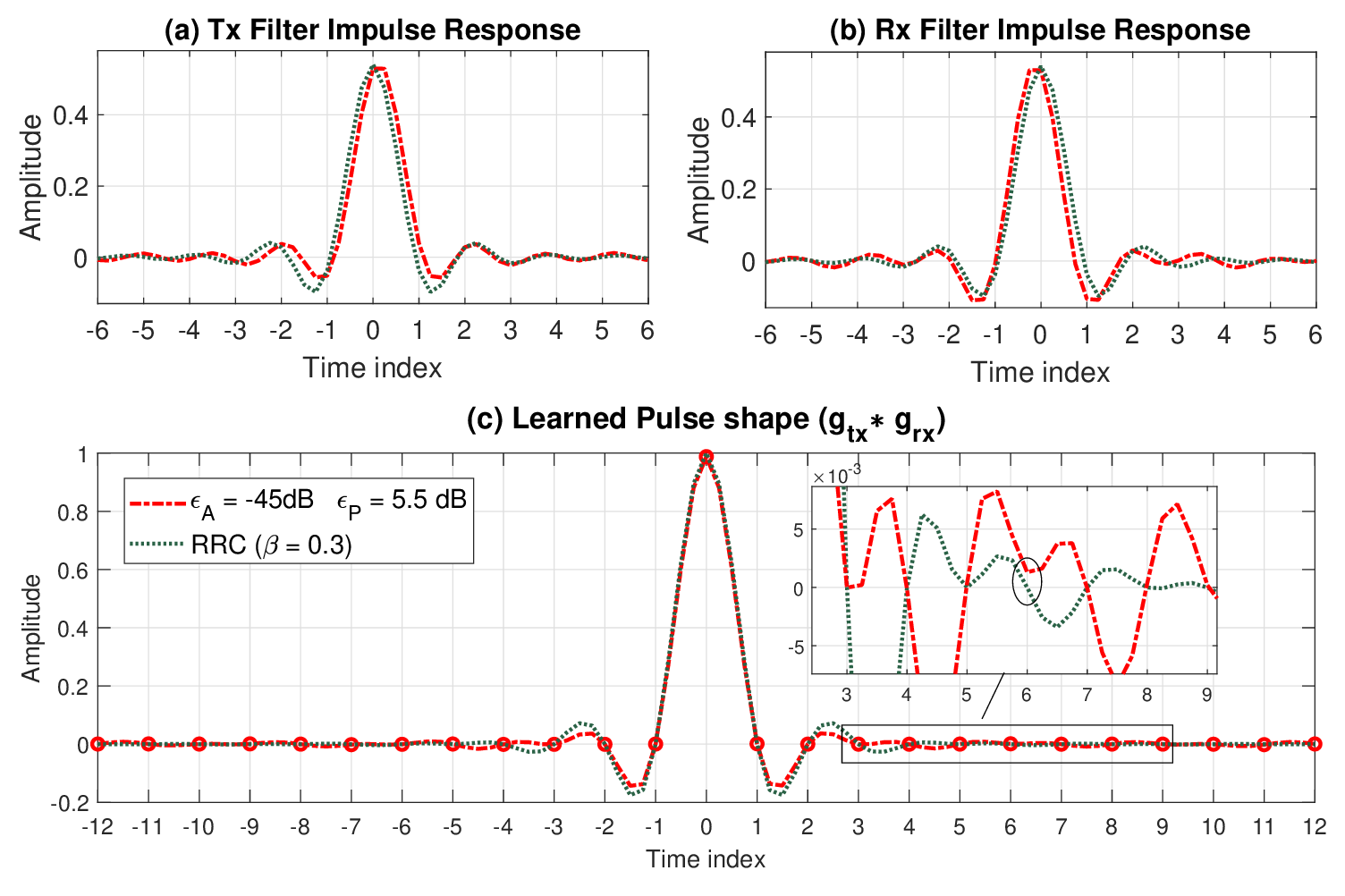}
    \caption{Learned filters and pulse shape for $\epsilon_P = 5.5$ dB at 120 GHz under $\epsilon_A = -45\text{ dB}$ with $\beta = 0.3$ with NND}
    \label{fig:learnt_filters}
\end{figure}

Now, \DMMinorRev{let us} delve into the characteristics of the learned filters in both the time and frequency domains. Figure \ref{fig:learnt_filters}a and \ref{fig:learnt_filters}b showcase the impulse responses of the learned Tx and Rx filters while their resultant pulse shape is shown in Figure \ref{fig:learnt_filters}c for a PAPR target of $\epsilon_P = 5.5\text{ dB}$, under an $\epsilon_A = -45\text{ dB}$ with $\beta = 0.3$ using an NND. The impulse responses exhibit notable symmetries around their peaks while the Tx filter feature a reduced amplitude compared to RRC in the first side lobe outside the main lobe, which contributes to lower PAPR \cite{NyquistM}. This reduction introduces a slight increase in ripples in the outer regions. 
The Rx filter impulse has the opposite effect, where the first trough has higher amplitude compared to the RRC case. Furthermore, the Tx filter has a slight shift in time, which again has the opposite effect in the Rx filter. Consequently, the convolution of these two filters, which forms the overall pulse shape, is symmetric around $t=0$, akin to conventional pulse shapes. The effect of the reduced first ripple and a slight increase in outer ripples become apparent in the overall pulse shape. Moreover, upon closer inspection of the enlarged portion in Figure \ref{fig:learnt_filters}c, a notable difference in tail characteristics to RRC is evident while at the sampling points, a slight deviation compared to the zero crossings of conventional pulses is observed.
This contributes to ISI in the received signal since the Nyquist criterion is not strictly followed.  However, as highlighted in \cite{NyquistM}, the Nyquist criterion, considered a hard constraint, may be unnecessary in actual design. The relaxation of this criterion brings advantages in stopband attenuation and PAPR characteristics. Furthermore, \DMMinorRev{it is} important to remember that, since our system is trained end-to-end, the ISI introduced by the learned filters are effectively mitigated by the receiver. The end-to-end training ensures that the entire transceiver is optimized to handle and compensate for such effects, contributing to robust performance. We observed slight numerical variations in the impulse response characteristics across demappers, indicating that the filter shape is primarily influenced by the PAPR and ACLR constraints, within the limits of the permitted excess bandwidth.
\begin{figure}
    \centering
    \includegraphics[width=\columnwidth]{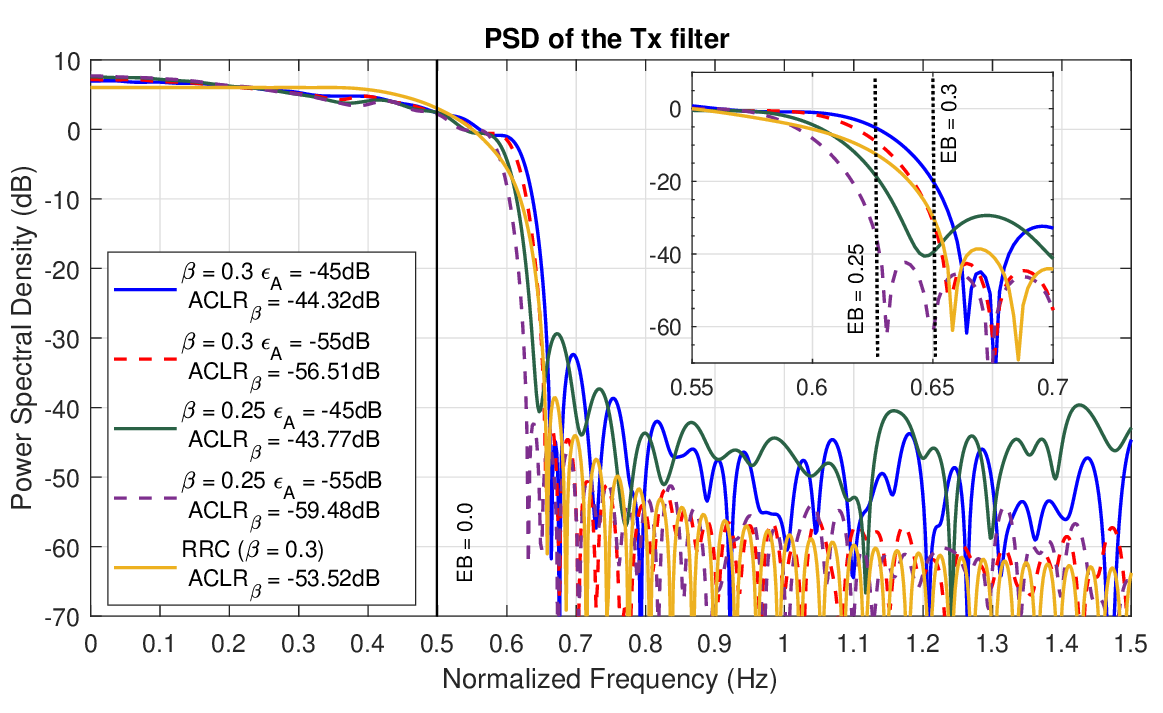}
    \caption{PSD of the Tx filters for different $\epsilon_A$ and $\beta$ at 120 GHz under $\epsilon_p =5.5\text{ dB}$ }
    \label{fig:PSD}
\end{figure}

Next, we examine the PSD of the learned Tx filters in Figure \ref{fig:PSD}. We present the PSD for four cases where, two ACLR$_\beta$ targets of $\epsilon_A = -45, -55\text{ dB}$ are considered for two excess bandwidths of $\beta = 0.25 , 0.3$, which are trained for $\epsilon_P = 5.5\text{ dB}$ under an NND. Note that RRC ($\beta = 0.3$) has an ACLR$_\beta$ of $-53.52 \text{ dB}$ outside the $1+\beta$ bandwidth. Although our $-45 \text{ dB}$ target is slighly higher compared to that, the stopband attenuation is sufficient for most practical cases.
One initial observation is that the learned filters have different stopband characteristics to RRC. Nevertheless, they approximately attain the specified ACLR$_\beta$ targets of $-45 \text{ dB}$ and  $-55 \text{ dB}$. Next, notice the roll-off characteristics in the transition region between the nominal bandwidth of 0.5 and 0.65 (0.15 excess bandwidth for one side). The learned filters exhibit higher PSD values in the passband with a slight ripple, followed by a steeper roll-off compared to the RRC. When the ACLR$_\beta$ target is lower at $\epsilon_A =-55\text{ dB}$, the roll-off starts earlier than the case where the ACLR$_\beta$ target is $-45 \text{ dB}$. Then the leakage outside the excess bandwidth is lower with $\epsilon_A =-55\text{ dB}$ and higher with $\epsilon_A =-45\text{ dB}$ compared to RRC. However, this is by design and negligible in these magnitudes. with the flexibility of our method, one could easily target a narrower excess bandwidth ($\beta < 0.3$), which will avoid the smaller leakage. We highlight the flexibility of the method to learn the filter for such a narrower excess bandwidth while still achieving the same ACLR targets and PAPR target, which presents an opportunity to enhance spectral efficiency, a demonstration of which will be presented later.

\subsection{BLER and SE performance} \label{BLERandSE}
Now that we have illustrated the constellation and filters and assessed the PAPR characteristics, we look at the performance of the waveforms with the considered setup. First we look at the baseline performance of QAM and APSK with an RRC under the AOD, PND-LPN and PND-HSNR. Then we select the best scheme for comparison with the learned waveforms under the three demappers at 120 GHz and 220 GHz. Furthermore, we demonstrate the flexibility of the method for learning under tighter excess bandwidth and showcase the advantage of the joint filter and constellation adaptation in comparison to our previous work.

\subsubsection{Baseline performances}
\begin{figure}[ht]
    \centering
    \includegraphics[width=\columnwidth]{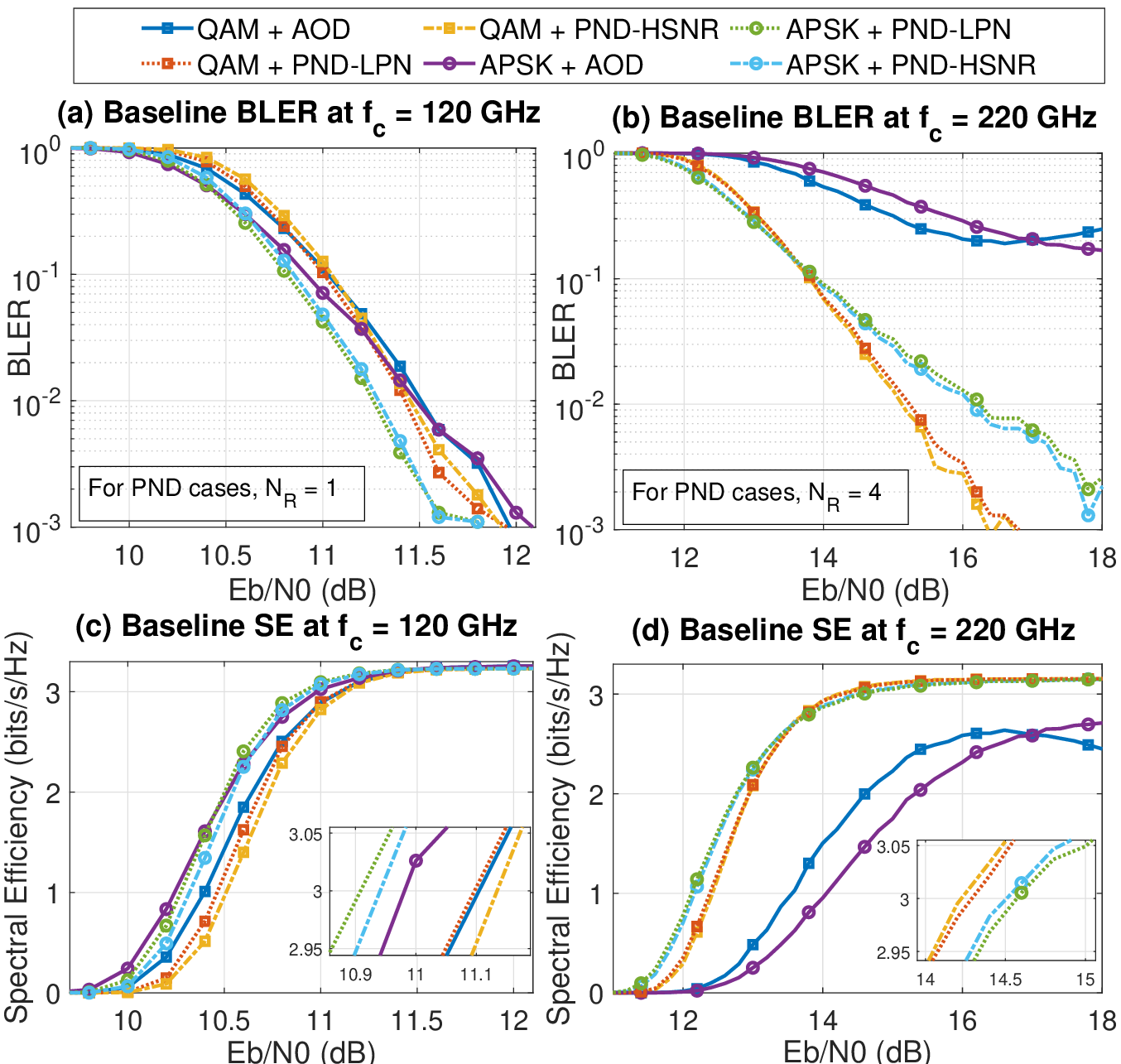}
    \caption{Baseline performance}
    \label{fig:Baseline_SE_BLER}
\end{figure}

First, \DMMinorRev{let us} look at the 120 GHz case, where the residual PN is low. Therefore, the needed number of total residual PN pilots is low for estimation of the PN variance. In this case, one additional pilot ($N_R=1$) has been appended to each PTRS block, which means a total of 32 residual pilots are utilized. We tested with an increased no. of residual PN pilots ($N_R=4$), which did not yield a noticeable improvement. Clearly from Figure \ref{fig:Baseline_SE_BLER}a, the APSK schemes perform better compared to QAM in all 3 demapping techniques. APSK coupled with PND schemes shows the best BLER performance, while PND-LPN has a slightly better performance. \DMRev{However, notice from Figure \ref{fig:Baseline_SE_BLER}b, that the achievable spectral efficiency of PND techniques are slightly lower compared to the AOD, due to the usage of the additional residual pilots.} Nevertheless, APSK+PND-LPN provides  the best baseline performance for our case at 120GHz. 

Now \DMMinorRev{let us} focus on the higher PN case at 220GHz, which is shown in Figures \ref{fig:Baseline_SE_BLER}b and \ref{fig:Baseline_SE_BLER}d. One immediate observation is that the AOD struggles to demap the received symbols properly, resulting in poor performance. 
We used $N_R = 4$ in the case of 220GHz, which totals to 128 pilots for PN variance estimation. With the PN variance estimated, PND schemes are able to demap the signal, while the PND-HSNR has slighlty better performance, in contrast to the 120GHz case. Comparing the constellations patterns, QAM performs significantly better compared to APSK in the higher  $E_b/N_0$ region, while APSK has a moderately better BLER in the lower $E_b/N_0$ region until the BLER reaches 10\%. However, recall that QAM has a higher PAPR compared to APSK (see Figure \ref{fig:CCDF}), which calls for a trade-off between the BLER, SE performance and PAPR. 

\subsubsection{Learned Waveforms in Low PN}
\begin{figure}[ht]
    \centering
    \includegraphics[width=\columnwidth]{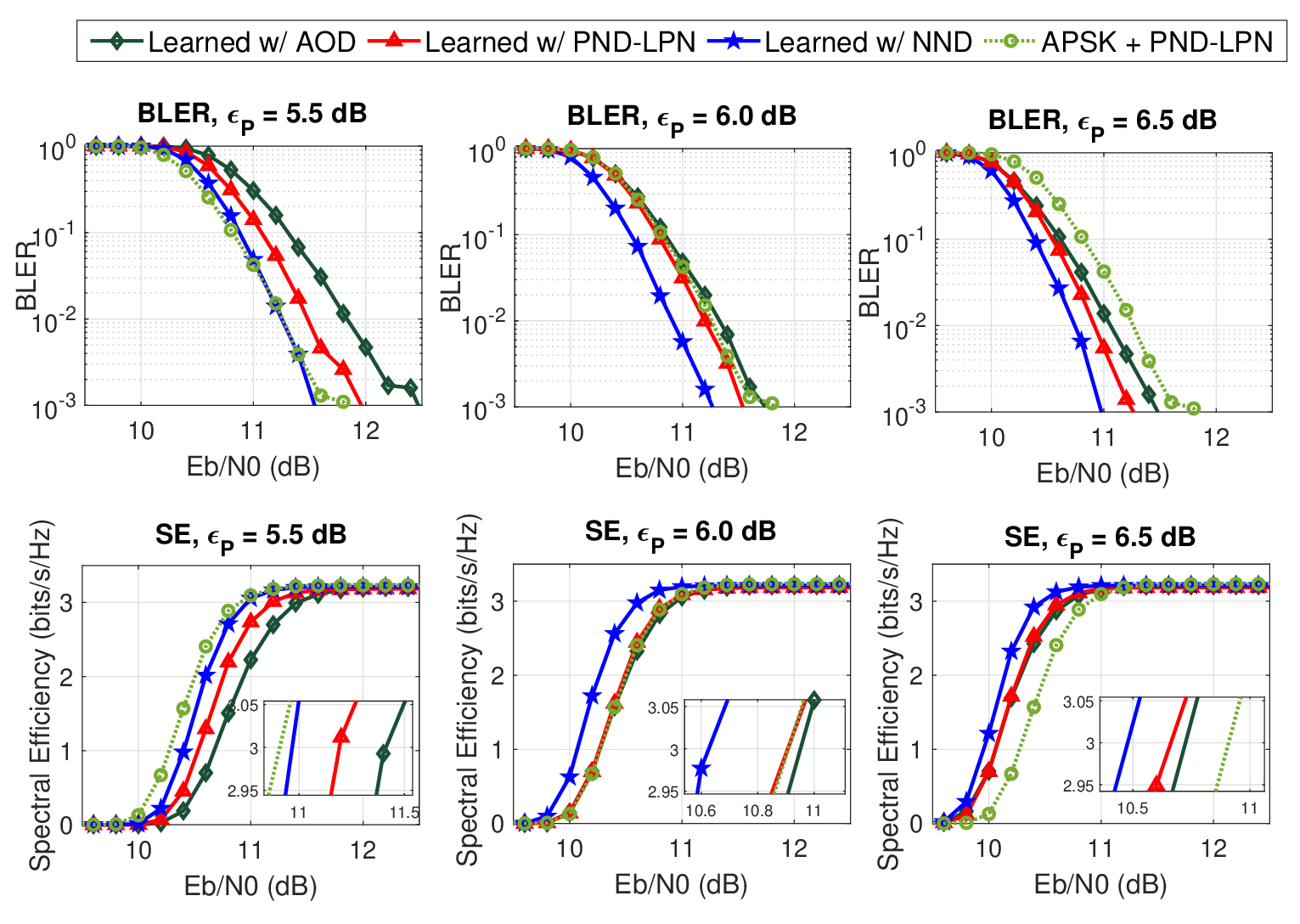}
    \caption{Performance of learned waveforms for 120 GHz $\beta =0.3$ $\epsilon_A = -45 \text{ dB}$ with baseline performance of APSK and PND-LPN}
    \label{fig:BLER_SE_120 GHz}
\end{figure}

Figure \ref{fig:BLER_SE_120 GHz} shows the BLER (on the upper row) and the corresponding SE (on the lower row) curves for the learned waveforms at 120 GHz with $\beta = 0.3$ and $\epsilon_A = -45\text{ dB}$, while the best baseline from Section \ref{BLERandSE}, APSK with PND-LPN is also shown for comparison. Each column corresponds to a target $\epsilon_P = 5.5, 6.0, 6.5 \text{ dB}$ respectively. 

With the stricter PAPR target at $\epsilon_P = 5.5\text{ dB}$, the NND delivers competitive BLER performance compared to the APSK+PND-LPN case. \DMRev{Note that, the learned waveform occupies a slightly wider bandwidth (refer to Figure \ref{fig:PSD}), which is reflected in a slight peak SE loss compared to the APSK+PND-LPN case.} However, this is outweighed by the 1.2 dB reduction in PAPR compared to waveforms employing APSK. \DMMinorRev{The other two learned waveforms suffer from the strict PAPR penalty, which have worse performance compared to the baseline.} This highlights the effectiveness of using the NND for improved adaptability, in contrast to the analytical demappers. Nevertheless, when the PAPR targets are relaxed to $6.0 \text{ dB}$ and $6.5 \text{ dB}$, it becomes evident that the learned waveforms under all three demapperrs deliver better BLER and SE performance. As expected, the NND performs the best, achieving nearly a $0.35 \text{ dB}$ gain in the required $E_b/N_0$ at 1\% BLER and SE at $\epsilon_P = 6.0\text{ dB}$. \DMMinorRev{It is} important to highlight that these improvements are achieved still with a reduction in PAPR of $0.8 \text{ dB}$ (refer to Figure \ref{fig:CCDF}). The BLER and SE gain margins increase with $\epsilon_P = 6.5\text{ dB}$, which has only a $0.3\text{ dB}$ gain in PAPR. \DMMinorRev{
In Table \ref{120_table}, we list the numerical performance values at $E_b/N_0 = 10.6\:dB$, where at least one waveform achieves 3 bits/s/Hz under $\epsilon_P = 6.5\:dB$, from the simulated points. This highlights the BLER and SE gain allowed by the learned waveforms, especially with the NND while achieving a similar PAPR performance.}

\begin{table}[ht] \caption{Numerical performance values for the case 120GHz, $\epsilon_P = 6.5\:dB, \epsilon_A = -45\: dB, \beta = 0.3,$ \\ @  $E_b/N_0 = 10.6\:dB$} \label{120_table}
\centering
\begin{tabular}{|cccc|}
\hline
\multicolumn{1}{|c|}{Scheme}                          & \multicolumn{1}{c|}{BLER}            & \multicolumn{1}{c|}{SE}           & SE Gain over BL       \\ \hline
\multicolumn{1}{|c|}{APSK + PND-LPN}            & \multicolumn{1}{c|}{$2.56 \times 10^{-1}$}        & \multicolumn{1}{c|}{2.41}       & -                     \\ \hline
\multicolumn{1}{|c|}{Learned w/ AOD}            & \multicolumn{1}{c|}{$1.06\times 10^{-1}$}        & \multicolumn{1}{c|}{2.87}       & 0.46                \\ \hline
\multicolumn{1}{|c|}{Learned w/ PND-LPN}        & \multicolumn{1}{c|}{$7.42\times 10^{-2}$}        & \multicolumn{1}{c|}{2.95}       & 0.54                \\ \hline
\multicolumn{1}{|c|}{Learned w/ NND}            & \multicolumn{1}{c|}{$2.72\times 10^{-2}$}        & \multicolumn{1}{c|}{3.13}       & 0.72                \\ \hline
\end{tabular}
\end{table}

\subsubsection{Learned Waveforms in High PN}

\begin{figure}[ht]
    \centering
    \includegraphics[width=\columnwidth]{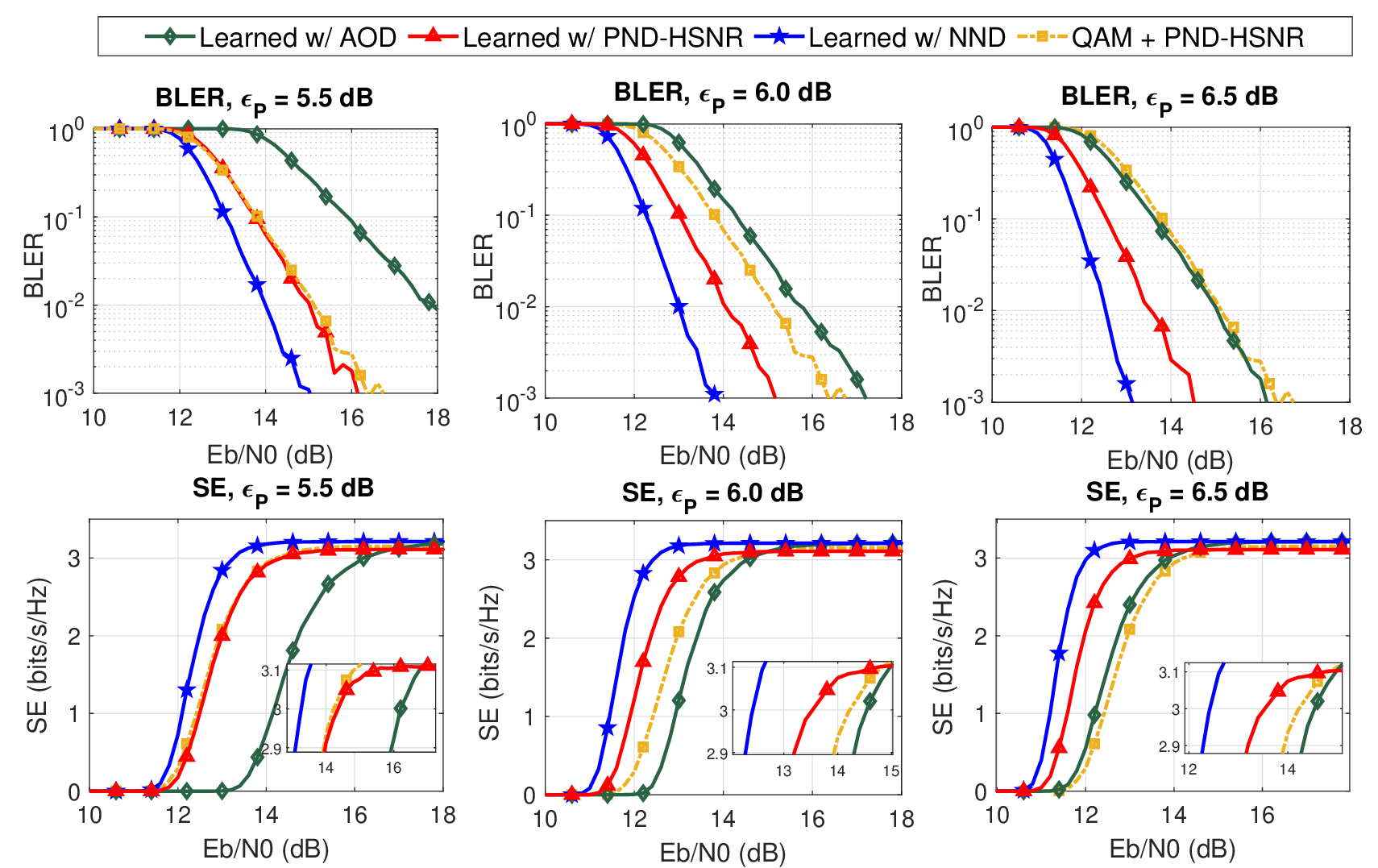}
    \caption{Performance of learned waveforms for 220 GHz $\beta =0.3$ $\epsilon_A = -45 \text{ dB}$ with baseline performance of QAM and PND-HSNR}
    \label{fig:BLER_SE_220 GHz}
\end{figure}

Next, \DMMinorRev{let us} shift our focus to the 220 GHz cases, demonstrated in Figure \ref{fig:BLER_SE_220 GHz}. The baseline shown is with the QAM+PND-HSNR case as identified in Section \ref{BLERandSE}, even though the PAPR of QAM is higher compared to APSK.  With the increase in frequency, the severity of residual PN intensifies, as evident from the learned constellations, especially under the AOD (refer to Figure \ref{fig:constellations}). Nevertheless, the waveforms learned with our method consistently deliver improved BLER and SE, except with the AOD, compared to their baseline counterparts. When compared to the PND-HSNR baseline, competitive performance of AOD is observed when the PAPR target is relaxed to $\epsilon_P = 6.5\text{ dB}$.  This can be attributed to the branching effect under the Euclidean metric, which makes the constellation points to be densely packed along the branches, leading to a higher SNR for demapping. Under both NND and PND, the learned waveforms consistently deliver superior BLER and SE in all the PAPR targets. At the lowest $\epsilon_P = 5.5\text{ dB}$, learned waveform under PND-HSNR matches the baseline. When the PAPR constraint is relaxed better BLER and SE can be observed. \DMRev{The use of NND in the higher PN clearly delivers the best performance, along with better SE since it does not need PN variance estimation. Specifically, notice that at $\epsilon_P = 6.5\text{ dB}$, a $2.5\text{ dB}$ gain in required $E_b/N_0$ at 1\% BLER compared to the baseline is achieved along with a reduction in PAPR of $0.5 \text{ dB}$, which is a commendable improvement under strong PN while the gain for PND is at $1.5\text{ dB}$ $E_b/N_0$}.
 \DMMinorRev{
Similar to earlier case, we list the numerical performance values at $E_b/N_0 = 13\:dB$ for 220 GHz in Table \ref{220_table}. Note that PND almost achieves an SE of 3 bits/s/Hz, while NND is clearly better. Most importantly, the gain over the best baseline is 1.12 bits/s/Hz, which highlights the superiority of the NND scheme especially when the PN is higher.}
\begin{table}[ht] \caption{Numerical performance values for the case 220GHz, $\epsilon_P = 6.5\:dB, \epsilon_A = -45\: dB, \beta = 0.3,$ \\ @  $Eb/N0 = 13 \:dB$} \label{220_table}
\centering
\begin{tabular}{|cccc|}
\hline
\multicolumn{1}{|c|}{Scheme}                                          & \multicolumn{1}{c|}{BLER}                            & \multicolumn{1}{c|}{SE}                           & SE Gain over BL                       \\ \hline
\multicolumn{1}{|c|}{QAM +   PND-HSNR}                          & \multicolumn{1}{c|}{$3.40\times10^{-1} $}                       & \multicolumn{1}{c|}{2.08}                       & -                                     \\ \hline
\multicolumn{1}{|c|}{Learned w/ AOD}                            & \multicolumn{1}{c|}{$2.53\times10^{-1} $}                       & \multicolumn{1}{c|}{2.34}                       & 0.32                                \\ \hline
\multicolumn{1}{|c|}{Learned w/ PND-HSNR}                        & \multicolumn{1}{c|}{$3.87\times10^{-2} $}                       & \multicolumn{1}{c|}{2.99}                       & 0.91                                \\ \hline
\multicolumn{1}{|c|}{Learned w/ NND}                            & \multicolumn{1}{c|}{$1.60\times10^{-3} $}                       & \multicolumn{1}{c|}{3.21}                       & 1.12                                \\ \hline
\end{tabular}
\end{table}

\subsubsection{Learned Waveforms with Lower Excess Bandwidth}

\begin{figure}[ht]
    \centering
    \includegraphics[width=\columnwidth]{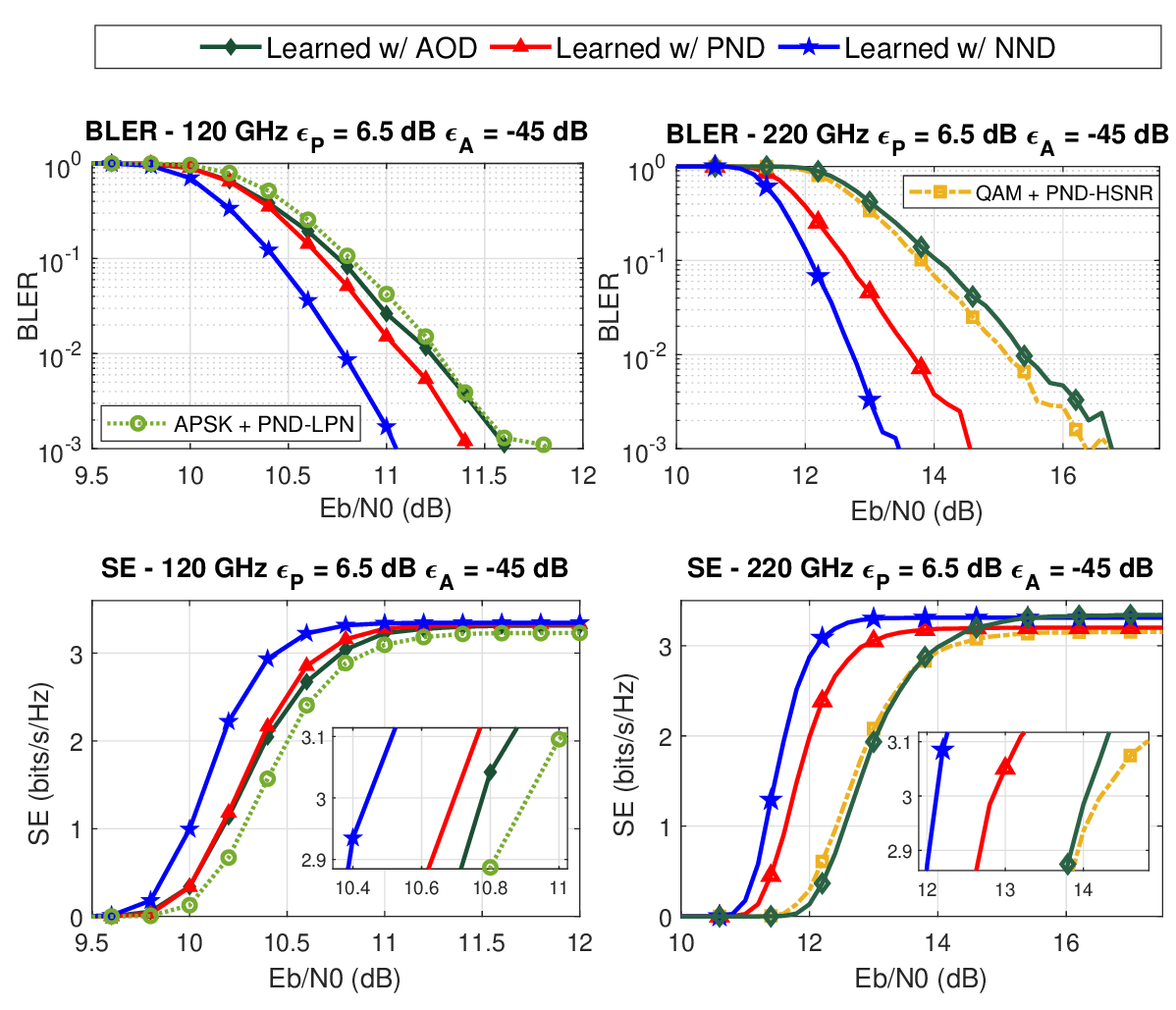}
    \caption{Performance of learned waveforms for $\beta =0.25$ and $\epsilon_A = -45\text{ dB}$ at 120 GHz and 220 GHz with respective baseline performances}
    \label{fig:BLER_SE_0_25}
\end{figure}

Now, we showcase the ability of our method to deliver better spectral efficiency by targeting a lower excess bandwidth of $\beta = 0.25$ compared to our baseline case of $\beta = 0.3$. In Figure \ref{fig:BLER_SE_0_25}, the BLER and SE performance of the learned waveforms at both 120 GHz and 220 GHz cases, with an $\epsilon_P = 6.5\text{ dB}$ are shown. The ACLR target for each waveform is $-45 \text{ dB}$.
In all cases except under AOD at 220 GHz, better BLER compared to the baseline is seen. 
\DMRev{The better BLER and the lower targeted excess bandwidth, hence implicitly lower OBW, leads to a significant gain in SE.} 

\subsubsection{Comparison with geometrical constellation shaping only}
\begin{figure}[ht]
    \centering
    \includegraphics[width=\columnwidth]{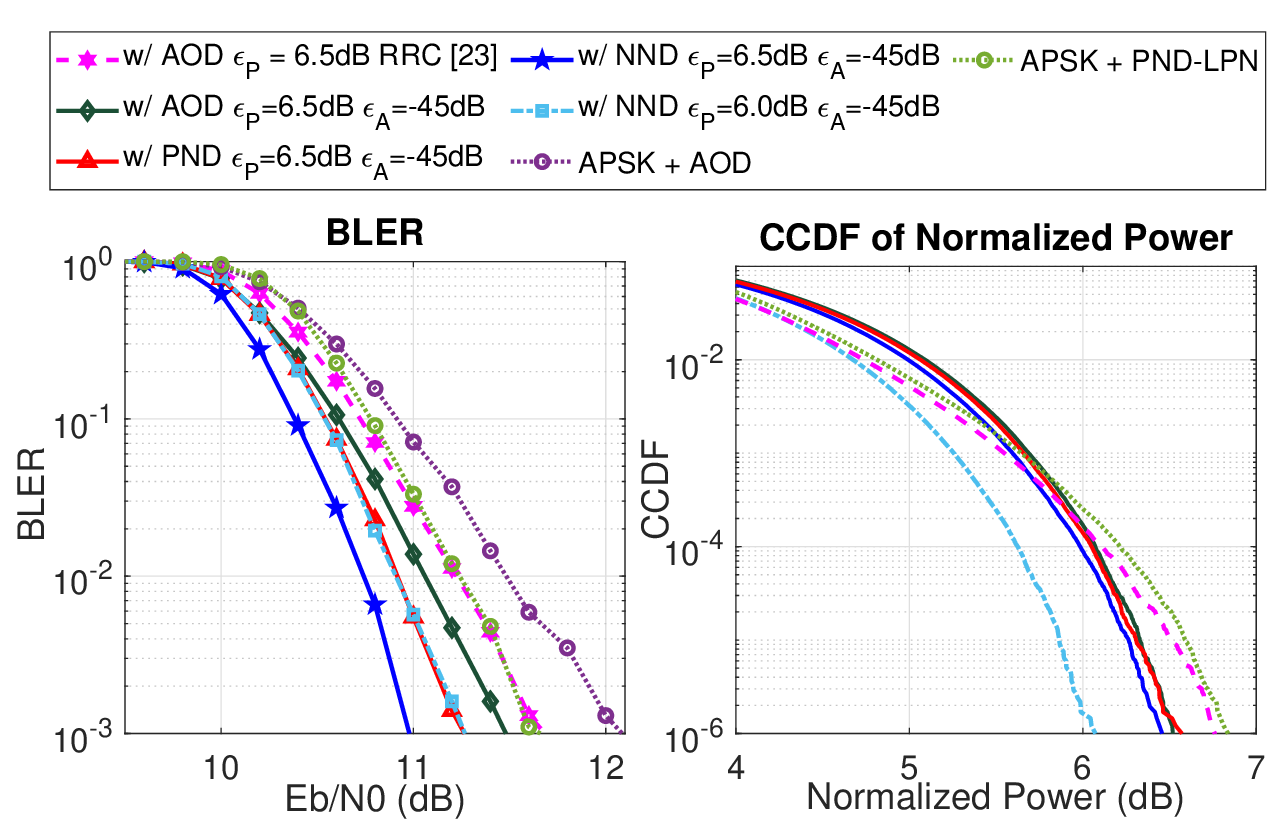}
    \caption{Comparison of constellation shaping \cite{ICCconstellation} and jointly optimization of constellation and filters. 
    }
    \label{fig:Conf_vs_journal}
\end{figure}

In our previous work\cite{ICCconstellation}, we presented constellation shaping under the PAPR constraint in low PN with an AOD, which does not have a trainable filter pair. Now, we compare the performance of this work, where the filter is incorporated in the end-to-end optimization under an ACLR constraint. Figure \ref{fig:Conf_vs_journal}, shows BLER and CCDF comparison, with the best performing scheme from \cite{ICCconstellation}, with  $\epsilon_P = 6.5\text{ dB}$. \DM{Also, the performance of APSK+AOD, which was the baseline in \cite{ICCconstellation} is shown for comparison.} The learned waveforms shown are subjected to an ACLR constraint $\epsilon_A = -45\text{ dB}$, with first 3 constrained with the same $\epsilon_P = 6.5\text{ dB}$ under the 3 demappers, while the fourth has a lower target of $\epsilon_P = 6.0\text{ dB}$. The two baselines shown are with APSK and RRC $\beta = 0.3$ under AOD and PND-LPN. First note that, all the learned waveforms have better PAPR compared to the constellation shaping case, even with the same $\epsilon_P$ target. When the BLER performance is considered, we can see the resulting constellation from \cite{ICCconstellation} performs similar to the APSK+PND-LPN baseline, while the former does not use any PN pilots, which means better spectral efficiency. 
The 4 learned waveforms outperform the result from \cite{ICCconstellation} in terms of BLER still possessing better PAPR characteristics. With AOD, the gain is $0.2 \text{ dB}$ at 1\% BLER, which is extended up to $0.5 \text{ dB}$ using the NND with same  $\epsilon_P = 6.5\text{ dB}$. With $\epsilon_P = 6.0\text{ dB}$ \DMMinorRev{for NND}, 
BLER gain reduces to the same level as PND, which is approximately $0.3 \text{ dB}$. This proves the fact that joint training of the constellation and filter pair leads to better performance. It further concretes the advantage of deploying NND.

\subsection{Discussion}
\DMMinorRev{The severity of the residual PN depends on the noise floors of the PSDs \LH{ and the utilized bandwidth}, and its impact varies with different modulation orders. The considered receiver PN model has a higher noise floor \LH{compared to the transmitter PN model} (see Fig. \ref{fig:PN_models}), hence contributing more to the overall system residual PN level. \LH{Considering only the receiver PN for our investigated sub-THz system of 3.93 GHz channel bandwidth, a spectral density of the Gaussian PN floor of 120~dBc/Hz at 120~GHz and 115~dBc/Hz at 220~GHz carrier frequency (see Fig. \ref{fig:PN_models}) leads to the contribution of an equivalent white Gaussian PN model with variance $\sigma_p^2 = 3.9 \times10^{-3}$ and $\sigma_p^2 = 1.24\times10^{-2}$, respectively, according to \cite{CEA_Leti_PN_model}. Following the Gaussian PN categorization therein, our investigated scenarios are in between low to medium PN cases. However, the modulation order has to be factored in, when determining the impact of the severity.}} \DMMinorRev{Furthermore, we measured the residual PN variances after Wiener PN compensation with 32x4 PTRS samples at $E_b/N_0 = 30 \: dB$ to isolate the Gaussian PN contribution, which was found to be $\sigma_p^2 = 3.2\times10^{-3}$ and $\sigma_p^2 = 1.08\times10^{-2}$, respectively. These measured variances are slightly lower than the calculated values based on \cite{CEA_Leti_PN_model}, most likely due to the PTRS-based compensation contributing to a lightly remedying of Gaussian PN as well.
 Since we use data-driven model-based methods, we utilized two specific PN models, but one could easily utilize the method for their own setups once, the local oscillator characteristics are known.} With the presented results, it is evident that the proposed method allows a flexible technique to design a sub-THz waveform under the PN impairment while targeting lower PAPR. Shaping the constellation and adapting the filter impulse response allowed the PAPR reduction achieved. The small trade-off in bandwidth can be avoided targeting a narrower excess bandwidth as discussed earlier. The main reason is that the roll-off of the learned waveforms \DM{is} different to that of RRC. The sharp roll-off allows one to efficiently use the allowed excess bandwidth and more fine tuning is possible by targeting different PAPR, ACLR and excess bandwidth targets as our framework allows such flexibility. The NND provides more adaptability to the PN condition, thus deliver better BLER and SE at all times compared to the baseline solutions. Note that, the NND does not require the noise variances as input, which frees up resources for data symbols unlike the analytical demappers. However, the NND performance is sensitive to the trained $E_b/N_0$ range.
This is a known limitation in NN based demappers and receivers, which might be alleviated with a technique such as SNR-deweighted training \cite{StuderBLER}. Each learned NND is specialized to the respective waveform learned, which is the combination of the constellation and filters while analytical demappers allow generalizability. Therefore, we showcased the feasibility of waveform optimization targeting the analytical demappers, the PND-LPN, PND-HSNR and AOD.

%% file: 8_ConculsiosnsAndOutlook.tex
This paper explores the optimization of a waveform for an \DM{SC} transceiver under phase noise impairment for sub-THz communications. The optimization is achieved through the shaping of constellations and adaptation of filters, while adhering to PAPR and ACLR constraints within a specified excess bandwidth. The solution approach was training in an end-to-end fashion under different demapping techniques at the receiver. The results proved the feasibility of learning a waveform to identify an optimized constellation geometry and pulse shaping filters while providing gains in terms of PAPR, BLER and SE compared to the conventional baselines. PAPR reductions up to 1.2 dB could be achieved utilizing an NN demapper, while delivering competitive BLER and SE performance. At the stronger PN conditions, gains up to $2.2\text{ dB}$ in required $E_b/N_0$ at 1\% BLER are achieved compared to conventional constellations and state-of-the-art PN demappers still delivering a PAPR gains up to $0.5 \text{ dB}$. Future work includes incorporating a \DM{non-linear} power amplifier and evaluating the SC-FDE transceiver under sub-THz channel models \DM{and considering multi-antenna systems.}